\documentclass[11pt,german]{article}
\columnwidth\textwidth

\usepackage{graphics} 
\usepackage{amsmath}
\usepackage{amssymb}
\usepackage{times}
\usepackage{dsfont}
\usepackage{cancel}
\usepackage[T1]{fontenc}
\usepackage[ngerman]{babel}

\usepackage{graphicx}

\begin{document}

\title{Die kontrollierte Kettenreaktion\\
(The controlled nuclear chain reaction)}
\author{Andreas Aste\\
Paul Scherrer Institute, CH-5232 Villigen PSI, Switzerland}
\date{September 29, 2015}
\maketitle 
\begin{abstract}
\begin{center}
\noindent Es werden die wichtigsten Aspekte der theoretischen Beschreibung einer kontrollierten nuklearen Kettenreaktion
in kompakter Form diskutiert.
\end{center}
\end{abstract}
\begin{center}
\end{center}

\subsection*{Einf\"uhrung}
Der Inhalt der vorliegenden Schrift befasst sich mit der quantitativen Beschreibung der Kettenreaktion,
wie sie in (kommerziellen) leichtwassermoderierten Leistungsreaktoren oder Forschungsreaktoren zu steuern ist.
Das eigentliche Prinzip der Kettenreaktion ist simpel: die Spaltung von $^{235}$U-, $^{239}$Pu-
oder $^{233}$U-Kernen durch moderierte, d.h. nieder\-ener\-ge\-tische Neutronen setzt wiederum
Neutronen frei, welche potenziell weitere Spaltungen im Brennstoff des Reaktors ausl\"osen.
Im Gegensatz zur unkontrollierten Kettenreaktion, bei welcher hochenergetische Neutronen
im komprimierten, hochangereicherten Spaltstoff einer Kernwaffe innerhalb k\"urzester Zeit ($10^{-6}$s)
einen grossen Anteil der vorhandenen Kerne spalten und deren Spaltenergie
freisetzen, ist die Kettenreaktion im Reaktor dank dem Vorhandensein verz\"ogerter Neutronen
steuerbar. Diese werden nicht wie die grosse Mehrzahl der Neutronen unmittelbar bei oder kurz nach der Kernspaltung
durch die Spaltfragmente innerhalb von etwa $10^{-14}$ Sekunden freigesetzt, sie werden
vielmehr mit einer durchschnittlichen Verz\"ogerung von etwa 13 Sekunden durch gewisse bei der Spaltung
entstehende Kernbruchst\"ucke emittiert, welche als Vorl\"aufer bezeichnet werden.
Ebenso wie die bei der Kernspaltung prompt freigesetzten
Neutronen m\"ussen auch die verz\"ogerten Neutronen durch ein Moderatormaterial abgebremst
werden, damit sie mit ausreichender Wahrscheinlichkeit weitere Kernspaltungen ausl\"osen k\"onnen -
schnelle Neutronen mit Geschwindigkeiten im Bereich von einigen tausend Kilometern pro Sekunde spalten
die Spaltkerne nicht sehr effektiv.
\\

\noindent Moderierte Neutronen bewegen sich in einem Reaktor bei Temperaturen um $300^o$C mit einer
charakteristischen Geschwindigkeit von etwa 3km/s. Auch wenn diese Neutronen fortw\"ahrend Kollisionen mit
Atomkernen erleiden und daher jeweils nur geringe Distanzen der Gr\"ossenordnung eines Zentimeters 
von Stoss zu Stoss zur\"ucklegen
k\"onnen, so breitet sich eine lokale St\"orung der Neutronendichte innerhalb eines Reaktors von wenigen Metern
r\"aumlicher Ausdehnung innerhalb Bruchteilen einer Sekunde aus. Dieses Verhalten rechtfertigt weitestgehend
eine kollektive Beschreibung der Neutronen durch das sogenannte \emph{Punktmodell}.
Ausgehend von einem Minimum an Annahmen und Daten k\"onnen im Rahmen dieses Modells zahlreiche
interessante Aussagen \"uber die Kinetik der Neutronenbilanz gemacht werden.

\subsection*{Station\"arer Reaktor (Punktmodell)}
Im Rahmen des sogenannten Punktmodells beschreibt man einen Reaktorkern als homogenes Objekt ohne
innere Substruktur - quasi als Punkt. Durch dieses Modell k\"onnen ausgehend von einigen grundlegenden
Beobachtungen weitgehende Konsequenzen abgeleitet werden,
welche in der Folge dargelegt werden sollen.
\\

\noindent In diesem Abschnitt betrachten wir vorerst den station\"aren Reaktor, in welchem gewisse physikalischen Gr\"ossen
wie beispielsweise die Zahl freier Neutronen im Reaktor oder die Leistung des Reaktors
zeitlich konstant sind - zumindest auf makroskopischen Zeitskalen der Gr\"ossenordnung Stunden oder gar l\"anger.
Als erste Annahme setzen wir an, dass die \emph{thermische} Neutronenproduktionrate $P$ in einem
Reaktor \emph{durch Kettenreaktion} proportional zur Anzahl bereits vorhandener thermischer
Neutronen ist:
\begin{equation}
P=\mathcal{P} \cdot n \, . \label{propor}
\end{equation}
Dies ist einleuchtend, da eine Verdoppelung der Neutronenzahl im Reaktor auch eine Verdoppelung der durch
die Neutronen induzierten Kernspaltungen nach sich ziehen w\"urde, welche wiederum die Quelle f\"ur die Frei\-setzung
weiterer Neutronen sind. Tats\"achlich st\"osst die in Gleichung (\ref{propor}) implizit getroffene Annahme,
dass $\mathcal{P}$ eine Proportionalit\"ats\emph{konstante} ist,
sp\"atestens dann an ihre Grenzen, wenn eine Erh\"ohung der Neutronenzahl
und die damit verbundene Leistungserh\"ohung des Reaktors die Kettenreaktion durch die Erhitzung des
Brennstoffs und des Moderators zu beinflussen beginnt. Bei niedrigen Leistungen im Promillebereich, im sogenannten
Anfahrbe\-reich eines Reaktors, ist $\mathcal{P}=konst.$ aber eine gerechtfertigte Annahme.\\

\noindent Die \emph{Produktionswahrscheinlichkeit} $\mathcal{P}$ hat die physikalische Dimension einer inversen Zeit, entsprechend definiert
man die sogenannte \emph{Generationszeit} (\emph{mean generation time})
\begin{equation}
\Lambda = \frac{1}{\mathcal{P}} \, .
\end{equation}
In vielen kommerziellen Reaktoren gilt in guter N\"aherung $\Lambda \simeq 10^{-5}\rm s \, \dots \, 10^{-4}$s.
Die thermische Neutronenproduktionsrate $P$ in Reaktoren, welche beispielsweise eine thermische Leistung im
Gigawattbereich aufweisen, betr\"agt einige $10^{20}$s$^{-1}$. Da ein einzelnes Neutron zugleich aber nur
etwa $10^{-4}$ Sekunden in einem solchen Reaktor \"uberlebt, finden sich im Reaktor insgesamt etwa
$n=P/\mathcal{P} \approx 10^{16}$ (freie) Neutronen. Verteilt auf das Gesamtvolumen des Reaktorkerns folgt
daraus eine Neutronendichte $n^{_\Box}$ von rund $10^8$ Neutronen pro Kubikzentimeter.
Damit bilden die freien Neutronen in einem Leistungsreaktor ein \"ausserst d\"unnes Gas,
wenn man bedenkt, dass ein Kubikzentimeter
Wasser bei Raumtemperatur und Normaldruck etwa $3.3 \cdot 10^{22}$ Wassermolek\"ule
enth\"alt.\\

\noindent In sogenannten \emph{schnellen Br\"utern} ist die Generationszeit  mit etwa $10^{-6\ldots-7}$s bedeutend
k\"urzer als in einem thermischen
Reaktor, da dort die Neutronen wenig abgebremst (moderiert) werden und bei einer Energie von etwa $10^5$eV
Kernspaltungen in hoch angereichertem plutoniumhaltigem Material ausl\"osen.
Die in der Atom-, Kern- und Teilchenphysik verwendete Energieeinheit Elektron(en)volt mit der Einheitenbezeichnung
eV betr\"agt dabei etwa $1.602 \cdot 10^{-19}$J. Thermische Neutronen in einem thermischen Reaktor hingegen haben
statistisch verteilte kinetische Energien im Bereich von etwa $0.025$eV bei der Raumtemperatur von $293$K $\sim 20^{o}$C
und von $0.05$eV in einem Leistungsreaktor bei einer etwa doppelt so grossen (Moderator-)Temperatur von $573$K $\sim300^o$C.
Mit etwa  $10^{-8}$s ist die Generationszeit in einer unkontrollierten Kettenreaktion, wie sie bei einer
auf Kernspaltung basierenden Explosion einer Kernwaffe ($" \!$Atombombe$"$) auftritt, noch einmal bedeutend k\"urzer.
In dieser von der kontrollierten Kettenreaktion in einem thermischen Reaktor klar zu unterscheidenden
Situation l\"ost jede Kernspaltung durch schnelle Neutronen etwa zwei weitere Kernspaltungen aus,
sodass innerhalb von etwa $10^{-6}$s eine Energie von mehreren hundert Billionen Joule freigesetzt
werden kann.
\\ 

\noindent Die Produktionsrate $P$ l\"asst sich leicht aus der thermischen Leistung eines Reaktors absch\"atzen.
Bei einer Kernspaltung im Reaktor wird durchschnittlich eine Reaktionsw\"arme von etwa 196MeV freigesetzt, also
$1.96 \cdot 10^8 \cdot 1.6 \cdot 10^{-19}$J = $3.14 \cdot 10^{-11}$J.
Bei 3 Gigawatt ($\rm 3 \cdot 10^9Js^{-1}$)
thermischer Leistung folgt, dass im Reaktor somit etwa $\rm 3 \cdot 10^{9}J / 3
 \cdot 10^{-11}J = 10^{20}$ Kernspaltungen pro Sekunde statt finden. Weiter werden bei einer Kernspaltung
durchschnittlich $\nu = 2.5$ Neutronen hochenergetische Neutronen erzeugt, welche aber anschliessend
in den thermischen Energiebereich abgebremst werden m\"ussen. Der Anteil der Neutronen, die diesen
Moderationsprozess erfolgreich \"uberdauern, wird durch die sogenannte Moderationsf\"ahigkeit $W_{th}$ des
Reaktors beschrieben, welche beispielsweise f\"ur Druck- oder Siedewasserreaktoren im Bereich von $W_{th} \simeq 0.8$ liegt.
Somit entstehen pro Kernspaltung letztlich im Mittel etwa 2 thermische Neutronen, von denen im Mittel eines
wieder eine Kernspaltung bewirken muss, wenn die Kettenreaktion auf konstanten Niveau beibehalten werden
soll. Die \"ubersch\"ussigen Neutronen m\"ussen im Reaktormaterial eingefangen werden oder durch Leckage aus dem
Reaktor austreten. Die Absorption der Neutronen kann insbesondere durch Steuerst\"abe gezielt beeinflusst werden.
In Druckwasserreaktoren wird dem K\"uhlwasser zur Reaktivit\"atsregelung auf l\"angerfristiger Zeitskala
zudem Bor in Form von Bors\"aure beigesetzt, da sich das im nat\"urlichen Bor
vorhandene Bor-Isotop $^{10}$B als starker Neutronenabsorber zur Reaktivit\"atsbindung eignet.
Im Siedewasserreaktor kann durch
\"andern der K\"uhlwassermenge, welche den Reaktorkern durchfliesst (Kernmassenstrom), auch die
Moderationsf\"ahigkeit des Reaktors gezielt beeinflusst werden. Unvermeidlicher Einfang von Neutronen findet
auch in Kernspaltungsprodukten wie zum Beispiel den Isotopen $^{135}$Xe oder $^{149}$Sm, dem Moderator
oder im Brennstoff selbst statt. Die den Brennstoff einschliessenden H\"ullrohre bestehen aus einleuchtenden
Gr\"unden aus korrosionsbest\"andigen (Zirconium-)Le\-gierungen hoher Neutronendurchl\"assigkeit wie beispielsweise
Zirkalloy.\\

\noindent Eine ungef\"ahre Berechnungsformel f\"ur die Neutronenproduktionswahrscheinlichkeit $\mathcal{P}$ findet sich im Anhang
hergeleitet. 
\\

\noindent Anschaulich gesprochen entspricht die Generationszeit $\Lambda$ derjenigen Zeit, in der eine gewisse
Anzahl be\-reits vorhandener thermischer Neutronen in einem Reaktor mit konstant gehaltener Struktur
durch Kernspaltung dieselbe Anzahl neuer thermischer Neutronen erzeugt. Nat\"urlich werden bei einer (typischerweise
durch thermische Neutronen ausgel\"osten) Kernspaltung
zuerst schnelle Neutronen mit einer durch\-schnitt\-lichen ki\-ne\-tischen Energie von etwa $\bar{E}=\rm 2MeV=3.2\cdot10^{-13}J$ 
und einer entsprechenden Geschwindigkeit $\sqrt{2 \bar{E}/m_n} \simeq 20000$km$\cdot$s$^{-1}$ freigesetzt,
wobei die Neutronenruhemasse durch $m_n=1.675 \cdot 10^{-27}$kg gegeben ist und relativistische Effekte
bei den diskutierten Geschwindigkeiten noch keine relevante Rolle spielen. Die schnellen Neutronen werden
aber innerhalb einer Zeitskala von einigen million\-stel Sekunden in den thermischen Bereich abgebremst.
Thermische Neutronen werden oft durch eine etwas willk\"urliche \emph{cutoff}-Bedingung an ihre kinetische
Energie der Art $E_{kin} < 5 \cdot k_B T_M$ definiert. Dabei ist $T_M$ die Moderatortemperatur
und $k_B=1.38\cdot 10^{-23} \rm JK^{-1}$ die Boltzmann-Konstante.
\\

\noindent Von den im Reaktor vorhandenen thermischen Neutronen gehen also einige verlustig, ohne jemals eine
Kern-spal\-tung auszul\"osen. Einerseits werden Neutronen im Reaktormaterial absorbiert, 
andererseits werden auch thermische Neutronen aus dem Reaktor hinausgestreut (Leckage).
F\"ur die Verlust- oder \emph{Destruktionsrate} $D$ setzt man mit Hilfe der \emph{Destruktionswahrscheinlichkeit} $\mathcal{D}$
ganz allgemein an:
\begin{equation}
D= \mathcal{D} \cdot n \, .
\end{equation}
$\tau=1/\mathcal{D}$ ist die mittlere Lebensdauer eines thermischen Neutrons im Reaktor, gemessen ab dem Zeitpunkt
seiner Enstehung als promptes oder verz\"ogert emittiertes Neutron. Stellen wir uns als Gedankenexperiment
vor, dass in einem Reaktor pl\"otzlich keine neuen Neutronen mehr erzeugt w\"urden, so w\"urde die Anzahl verbleibender
Neutronen gem\"ass der Zerfallsgleichung
\begin{equation}
\frac{dn}{dt}=- \mathcal{D} \cdot n \quad \rightarrow \quad n(t)=n_0 \cdot e^{-\mathcal{D} \cdot t} =
n_0 \cdot e^{- t/\tau}
\end{equation}
abnehmen. Nach der Zeit $\tau$ w\"are die Anzahl der Neutronen um einen Faktor $e\simeq 2.718...$ reduziert,
und die mittlere Lebensdauer der Neutronen ist, wie aus der Theorie der radioaktiven Zerf\"alle bekannt, ebenso
durch $\tau$ gegeben. Die Destruktionswahrscheinlichkeit $\mathcal{D}=\mathcal{A}+\mathcal{L}$ l\"asst sich als Summe
einer Absorptionswahrscheinlichkeit $\mathcal{A}$ und einer Leckagewahrscheinlichkeit $\mathcal{L}$ darstellen.
\\

\noindent Schliesslich werden in einem Reaktor durch Zerf\"alle radioaktiver Substanzen fortlaufend Neutronen freigesetzt,
wobei es sich bei den betreffenden Zerf\"allen prim\"ar um \emph{spontane Spaltung} handelt.
Die spontan spaltenden Substanzen entstehen beim Betrieb des Reaktors (wichtig sind die
Curium-Isotope $^{242}$Cm und $^{244}$Cm);
zu\-gleich besteht die M\"oglichkeit, in einen Reaktor eine k\"unstliche Neutronenquelle gezielt einzubringen
(wie zum Beispiel eine $^{252}$Cf-Quelle, deren spontan spaltendes Californium in einem Hochflussreaktor
erbr\"utet wurde, oder eine sogenannte Alpha-Beryllium-Neutronenquelle oder eine
Gamma-Beryllium-Neutronenquelle). Die totale thermische Neutronenproduktionsrate aller
vorhandenen Quellen sei im Folgenden mit $Q_{th}$ bezeichnet.
\\

\noindent In einem station\"aren Kernreaktor ist die die \"Anderungsrate der Neutronenzahl $\dot{n}(t)=\frac{d}{dt}n(t)$ somit gegeben durch
\begin{equation}
\dot{n} = P-D+Q_{th} = (\mathcal{P}-\mathcal{D}) \cdot  n +Q_{th} = 0 \, . \label{stationaer}
\end{equation}

\noindent Weiter soll daher vorerst f\"ur unsere Betrachtungen $\mathcal{P} < \mathcal{D}$ gelten.
Der Reaktor ist dann unterkritisch, die Kettenreaktion
defizit\"ar, da durch diese mehr Neutronen vernichtet werden als erzeugt.
Dann kann Gleichung (\ref{stationaer}) in der Form
\begin{equation}
\frac{\mathcal{P}-\mathcal{D}}{\mathcal{P}} \cdot n + \Lambda \cdot Q_{th}=0
\end{equation}
geschrieben werden.  F\"uhrt man die praktische Abk\"urzung
\begin{equation}
\rho = \frac{\mathcal{P}-\mathcal{D}}{\mathcal{P}}
\end{equation}
ein, so gilt
\begin{equation}
\rho \cdot n + \Lambda \cdot Q_{th}=0 \, . \label{unterkritisch}
\end{equation}

\noindent Tats\"achlich l\"asst sich die Generationszeit $\Lambda$ in einem Reaktor ja n\"aherungsweise mit der mittleren Lebensdauer
$\tau$ vergleichen, die einem Neutron f\"ur das Durchlaufen des \emph{Neutronenzyklus} (welchen wir in der Folge
auch als \emph{Fermizyklus} bezeichnen wollen)
von seiner Entstehung bei einer Kernspaltung oder seiner Freisetzung durch einen Vorl\"aufer
bis zum Ausl\"osen einer erneuten Kernspaltung oder seiner Absorption
innerhalb oder ausserhalb des Reaktors zur Verf\"ugung steht. Pr\"aziser ausgedr\"uckt ist $\Lambda$ die Zeitspanne, in welcher sich
in einem station\"aren Reaktor eine Neutronenpopulation vollst\"andig selbst reproduziert. Im unterkritischen Reaktor
ist dabei $\Lambda > \tau$, offensichtlich sterben die Neutronen schneller als sie f\"ur ihre arterhaltende Fortpflanzung
brauchen w\"urden.
Gleichung (\ref{unterkritisch}) erh\"alt damit eine einfache Interpretation:
W\"ahrend einer Gene\-rations\-zeit $\Lambda$ werden  in einem Kernreaktor $\Lambda \cdot Q_{th}$ Neutronen
durch die Quelle(n) erzeugt, w\"ahrend von den vorhandenen $n$ Neutronen $\rho \cdot n$ Neutronen durch
den defizit\"aren Fermizyklus vernichtet werden. Im station\"aren Zustand halten sich diese beiden Prozesse
die Waage, und die Neutronenquellst\"arke des Reaktors gleicht den Neutronenverlust gerade aus.
\\

\noindent Die Gr\"osse $\rho$ wird \emph{Reaktivit\"at} genannt. Offensichtlich ist $\rho$ negativ, wenn $\mathcal{P}
< \mathcal{D}$. Im station\"aren  Fall mit $\rho<0$ (im sogenannten \emph{Quellbereich}) folgt aus Gleichung (\ref{unterkritisch}) 
\begin{equation}
n= -\frac{1}{\rho} \cdot  \Lambda \cdot Q_{th} \,. \label{invn}
\end{equation}

\noindent Da die Reaktivit\"at in der Praxis meist eine kleine Zahl $|\rho| \ll 1$ ist, wird sie \"ublicherweise in der Einheit
pcm (Abk\"urzung f\"ur \emph{pour cent mille, per cent mille}) angegeben.
Dabei ist 1pcm = $10^{-5}$, eine Reaktivit\"at von 0.0005 ist also gleich 50pcm.
\\

\noindent Ein Reaktor mit $\rho < 0$ heisst unterkritisch, und aus Gleichung (\ref{invn}) wird ersichtlich, dass die station\"are
Neutronenzahl in einem Reaktor stark ansteigt, wenn die  negative Reaktivit\"at eines Reaktors in die N\"ahe des
Wertes $\rho=0$ ger\"uckt wird. An diesem Anstieg von $n \sim -1/\rho$ l\"asst sich messtechnisch erkennen, wann der
kritische Zustand $\rho=0$ (oder $\mathcal{P}=\mathcal{D}$) eines Reaktor unmittelbar bevorsteht.
Das entsprechende \emph{kri\-ti\-sche Experiment} erm\"oglicht erst das sichere Kritischfahren eines Reaktors aus dem
Quellbereich heraus.
Der Quellbereich wird auch als Impulsbereich bezeichnet, da sich in einem stark unterkritischen Leistungsreaktor
($\rho < -1\% =-1000$pcm) oft nur noch wenige Millionen Neutronen befinden, die in den sich ausserhalb des Reaktors
befindenden Neutronendetektoren nur noch wenige Messimpulse pro Sekunde ausl\"osen, ganz im Gegensatz zum Leistungsbetrieb,
wo die Neutronenzahl m\"uhelos um zehn Zehnerpotenzen gr\"osser sein kann.
\\

\noindent Technisch kann die Reaktivit\"at beispielweise durch das \emph{Ausfahren} von Steuerst\"aben gesteuert
werden. Dadurch
werden im Reaktor weniger Neutronen absorbiert, die Destruktionsrate so gesenkt und $\rho$ erh\"oht.
Die resultierende Erh\"ohung der Neutronenzahl im Reaktor f\"uhrt zur Erh\"ohung der lokalen thermischen
Neutronenflussdichten, welche durch (externe) Neutronendetektoren detektiert werden k\"onnen.
Die Neutronenflussdichte $\Phi_{th}=n_{th}^{_\Box} \cdot \bar{v}$  ist definiert als das Produkt der thermischen
Neutronendichte $n_{th}^{_\Box}$ mit der (mittleren) Geschwindigkeit $\bar{v}$ der thermischen Neutronen.
Sie ist eine effektive Gr\"osse, welche mit der thermischen Reaktorleistung in unmittelbarem Zusammenhang steht.
Die oben verwendete explizite Indizierung der thermischen Neutronenzahl oder der thermischen Neutronendichte durch ${th}$
wird in dieser Arbeit aber nur im Bedarfsfall angewendet, da bei den meisten Betrachtungen sowieso nur thermische
Neutronen von Bedeutung sind.

\subsubsection*{{\emph{Der}} Multiplikationsfaktor}
Alternativ zur Reaktivit\"at $\rho$ ist auch die alternative formale Verwendung des Multiplikationsfaktors
\begin{equation}
k=\mathcal{P}/\mathcal{D}= \tau/\Lambda \label{multifac}
\end{equation}
von praktischem Nutzen. Gem\"ass der Definition besteht also die Beziehung
\begin{equation}
\rho=\frac{\mathcal{P}-\mathcal{D}}{\mathcal{P}}=\frac{k-1}{k}, \quad k=\frac{1}{1-\rho} \, .
\end{equation}
Damit kann Gleichung (\ref{unterkritisch}) auch in der Form
\begin{equation}
(k-1) n = -k \cdot\Lambda \cdot Q_{th}=-\frac{\mathcal{P}}{\mathcal{D}} \cdot \frac{1}{\mathcal{P}} \cdot Q_{th} = -\tau \cdot Q_{th}
\end{equation}
geschrieben werden. Innerhalb der mittleren Neutronenlebensdauer $\tau$ n\"ahme also die Neutronenzahl in einem station\"aren unterkritischen
Reaktor (mit $k<1$) von $n$ um den Multiplikationsfaktor $k$ auf $k \cdot n$ ab, w\"urde die Quelle den Neutronenverlust
$(k-1)n$ nicht mit $\tau \cdot Q_{th}$ Neutronen kompensieren. Im \"uberkritischen Reaktor mit $k>1$ w\"achst die
Neutronenzahl auch ohne Quelle $" \!$multiplikativ$"$ oder exponentiell an. Dieser dynamische Vorgang muss zum
besseren Verst\"andnis
aber einer etwas detaillierteren Betrachtung unterzogen werden.

\subsubsection*{Kinetische Grundgleichungen}
Wollten wir das instation\"are, zeitliche Verhalten der Neutronenzahl in einem Reaktor untersuchen, so w\"are ein voreiliger
Ansatz gegeben durch
\begin{equation}
\dot{n}= \frac{\rho}{\Lambda} \cdot n + Q_{th} \quad \mbox{oder} \quad     
\Lambda \cdot \dot{n}= {\rho} \cdot   n + \Lambda \cdot Q_{th} \, .
\end{equation}

\noindent Diese Gleichung dr\"uckt folgenden Umstand aus: W\"ahrend einer  Generationszeit werden im Kernreaktor
$\Lambda \cdot \dot{n}$ thermische Neutronen gebildet ($\dot{n}$ kann in dieser kurzen Zeitspanne als konstant betrachtet werden),
davon werden $\Lambda \cdot Q_{th}$ Neutronen durch die Quelle und $\rho \cdot n$ Neutronen durch die Kettenreaktion erzeugt.
\\

\noindent Nun werden bei einer Kernspaltung tats\"achlich nicht alle Neutronen sofort, also \emph{prompt} freigesetzt.
Es kommt vor, dass bei einer Kernspaltung Bruchst\"ucke des urspr\"unglichen Spaltkerns entstehen, die erst mit einer
zeitlichen Verz\"ogerung in der Gr\"ossenordnung von einigen Sekunden bis Minuten schliesslich doch noch
ein Neutron emittieren. Bereits 1939 wurde dieser Umstand bemerkt \cite{Roberts}.
Solche Kerne werden \emph{Vorl\"aufer} (\emph{delayed neutron precursors}) genannt. Bis 2001 waren
382 verschiedene Vorl\"aufer bekannt \cite{Pfeiffer}.
\\

\noindent In einem vereinfachten, aber zweckm\"assigen  Modell kann angenommen werden, dass alle Vorl\"aufer dieselbe
Lebensdauer besitzen. F\"ur viele Betrachtungen ist eine durchschnittliche Lebensdauer von $l=13$s
ein prak\-ti\-scher Wert.
Dann ist es so, dass in einem Reaktor fortlaufend Vorl\"aufer erzeugt werden, die
wiederum Neutronen freisetzen. Experimentell findet man, dass bei einer Kernspaltung in reinem $^{235}$U
etwa $\beta=\beta_{U-235}=0.64\%$ der Neutronen verz\"ogert freigesetzt werden.
Der Wert von 0.64\% f\"ur den Anteil verz\"ogerter Neutronen, welcher auf eine klassische Arbeit
\cite{Keepin} zur\"uckgeht und in der Literatur weit verbreitet ist, soll in der Folge als Grundlage verwendet werden.
Modernere Arbeiten favorisieren etwas h\"ohere Werte im Bereich $\beta_{U-235}=(0.665\pm 0.021)\%$
(siehe auch \cite{Piksaikin}).
\\

\noindent Die resultierenden Bilanzgleichungen f\"ur die Neutronenzahl $n$  und Vorl\"auferzahl $C$ lauten somit
\begin{equation}
\dot{n}  =  \frac{\rho - \beta}{\Lambda} n   +  \lambda C + Q_{th}  \label{gl1a}
\end{equation}
und
\begin{equation}
 \dot{C}  =  \frac{\beta}{\Lambda} n    -   \lambda C  \, . \label{gl2a}
\end{equation}
Diese Gleichungen werden auch (vereinfachte) \emph{kinetische Grundgleichungen} der Reaktorkinetik  oder 
reaktordynamische Punktgleichungen genannt.
\\

\noindent Der Term $-\frac{\beta}{\Lambda} \cdot n$ in der ersten Gleichung (\ref{gl1a}) ber\"ucksichtigt den Umstand, dass
in jedem einmal durchlaufenen neu\-tronen\-zahl\-er\-hal\-ten\-den Fermizyklus der Dauer
$\Lambda$ jeweils $\beta \cdot n$ Neutronen f\"ur die Vorl\"auferproduktion
abgezweigt werden. Zugleich zerfallen $C$ momentan vorhandene Vorl\"aufer mit einer Rate von
$\lambda \cdot C$ und setzen dabei ein Neutron frei. Die Zerfallsrate beim radioaktiven Zerfallsgesetz
ist \"uber die Beziehung $\lambda = 1/l$ mit der mittleren Lebensdauer der zerfallenden Kerne verkn\"upft.
\\

\noindent Die zweite Gleichung (\ref{gl2a}) beschreibt die Produktion der Vorl\"aufer durch Neutronen
($\beta \cdot n$ Vorl\"aufer pro Fermizyklus der Dauer $\Lambda$) und ber\"ucksichtigt auch deren Zerfallsrate durch den Term
$-\lambda C$.
\\

\noindent Tats\"achlich werden die prompten Neutronen mit einer durchschnittlichen Energie von etwa 2 MeV
(siehe Anhang), die ver\-z\"o\-ger\-ten Neutronen mit durchschnittlich etwa 0.5 MeV freigesetzt. Daher ist die
Ther\-ma\-li\-sierungs\-wahr\-schein\-lich\-keit,
also die Wahrscheinlichkeit, dass ein hochenergetisches Neutron in den thermi\-schen Bereich abgebremst werden kann,
f\"ur prompte und verz\"ogerte Neutronen nicht wirklich gleich. Die Gr\"ossen $C$ und $\beta$ in den ki\-ne\-ti\-schen
Grund\-glei\-chun\-gen entsprechen effektiven, thermisch gewichteten Gr\"ossen, die aber von der tats\"achlichen
Vorl\"auferzahl und dem naiv definierten $\beta$-Wert in realistischen F\"allen nicht stark abweichen.
Eine \"ahnliche Aussage gilt f\"ur die Neutronenzahl $n$ in den reaktorkinetischen Grundgleichungen (Punktgleichungen).
Ein Neutron, welches am Rande eines Reaktors entsteht, l\"ost auf Grund seiner hohen Lecka\-ge\-wahr\-schein\-lich\-keit mit geringerer
Wahrscheinlichkeit eine weitere Kernspaltung aus und tr\"agt so weniger zur Neutronenbilanz bei als ein im Inneren des
Reaktors entstandenes Neutron. Diesem Umstand kann mit Hilfe einer Gewichtungsfunktion oder
Einflussfunktion $\Phi^\dagger$ (\emph{importance function}) Rechnung getragen werden, aus welcher sich eine
f\"ur die Punktgleichungen relevante gewichtete Neutronenzahl der Form
\begin{equation}
n = \! \! \! \!  \int \limits_{Reaktorkern} \! \! \! \!  \! \! \! \! \! \!  \, \Phi^\dagger(\vec{r}) \,  n_{th}^{_\Box} (\vec{r}) \, dV
\end{equation}
errechnen l\"asst. 
Trotzdem ist es f\"ur das Verst\"andnis der Reaktorkinetik
hinreichend, wenn $n$ als eigentliche Gesamtzahl der thermischen Neutronen aufgefasst wird.
In guter N\"aherung kann $\Phi^\dagger$ zur im Reaktor herrschen Neutronenflussdichte $\Phi_{th}$
proportional gesetzt werden.
\\

\noindent Weiter muss bemerkt werden, dass die bei der Kernspaltung entstehenden Vorl\"auferkerne bei etwas
pr\"aziseren Betrachtungen typischerweise in sechs Gruppen von Vorl\"aufern mit jeweils vergleichbaren Lebensdauern
zu\-sammengefasst werden. Es wird in diesem Falle die Anzahl der in Gruppe $i$ zusammengefassten Vorl\"aufer,
welche ein verz\"ogertes Neutron
freisetzen werden, durch eine effektive Anzahl $C_i$ und eine mittlere Gruppen\-zerfalls\-kon\-stan\-te $\lambda_i$ angegeben.
Eine eigene "Gruppe" bildet der langlebigste Vorl\"aufer $^{87}$Br mit einer Lebensdauer von $1/\lambda_6 = 80.387$ Sekunden
oder einer Halbwertszeit von 55.72s (siehe Anhang). Ein $^{87}$Br-Kern wandelt sich aber mit 97\%-iger Wahrscheinlichkeit
durch $\beta^-$-Zerfall in einen $^{87}$Kr-Kern um, bei lediglich 2.6\% der Zerf\"alle vollzieht sich ein
$\beta^- n$-Zerfall, bei dem bei einem $\beta$-Zerfall zugleich noch ein Neutron emittiert wird - bei diesem Prozess entsteht
stabiles $^{86}$Kr. Die entsprechende Vorl\"auferzahl $C_6$ z\"ahlt nur diesen Anteil der f\"ur die Kettenreaktion relevanten $^{87}$Br-Kerne.
\\

\noindent Im station\"aren Reaktor gilt $\dot{C}=0$, wegen Gleichung (\ref{gl2a}) folgt
\begin{equation}
\dot{C}  =  0 = \frac{\beta}{\Lambda} n    -   \lambda C  \,  , \quad \mbox{es ist also} \quad 
C = \frac{\beta}{\lambda \Lambda} n
\simeq 10^3 n \, . \label{verhaeltnis}
\end{equation}
Die Vorl\"auferzahl \"ubertrifft die Neutronenzahl im station\"aren Reaktor bei Weitem und die
Vorl\"aufer stellen ein grosses, tr\"ages Reservoir an  latenten Neutronen dar.
\\

\noindent Offensichtlich ist im station\"aren Zustand auch $\lambda C = \frac{\beta}{\Lambda} n$ und somit folgt weiter mit
Gleichung (\ref{gl1a}) im station\"aren Zustand
\begin{equation}
0=\frac{\rho-\beta}{\Lambda} n + \lambda C + Q_{th} = \frac{\rho-\beta}{\Lambda} n + \frac{\beta}{\Lambda} n
+ Q_{th} = \frac{\rho}{\Lambda} n + Q_{th} \, , 
\end{equation}
wie es in den einf\"uhrenden Betrachtungen dargelegt wurde.

%------------------------------------------------------------------------------------------------------------------------------------

\subsection*{Zeitverhalten der Neutronenzahl im  Reaktor mit konstanter Reaktivit\"at $\rho$}
\subsubsection*{L\"osung der vereinfachten kinetischen Grundgleichungen: Vorbetrachtungen}

 \noindent Die vereinfachten kinetischen Grundgleichungen mit einer Vorl\"aufergruppe lauten f\"ur die (gewichtete)
Neutronenzahl $n$ und die thermisch gewichtete Vorl\"auferzahl $C$ bei Vernachl\"assigung der Quelle
\begin{equation}
\dot{n}  =  \frac{\rho - \beta}{\Lambda} n  \, \, +  \lambda C  \, , \label{gl1}
\end{equation}
\begin{equation}
\ \dot{C}  =  \frac{\beta}{\Lambda} n   -  \lambda C \, , \label{gl2}
\end{equation}
wobei $\Lambda \simeq 10^{-4}$ s die Generationszeit, $\beta \simeq 0.64\% = 0.0064$
den (effektiven) Anteil verz\"ogerter (thermischer) Neutronen in einem mit reinem $^{235}$U betriebenen
Reaktor und $\lambda = 1/l \simeq 0.0773$s$^{-1}$
die durchschnittliche Zerfallskonstante der Vorl\"aufer darstellt. Die Reaktivit\"at $\rho$ kann durch externe Einfl\"usse
kurzfristig variiert werden. F\"ur grobe Absch\"atzungen  kann  auch $\Lambda \simeq 10^{-4}$s,
$\lambda = 0.1$s$^{-1}$ oder $l = 10$s und $\beta=0.01= 10^{-2}$ angenommen werden.
\\

\noindent Bei schwach angereichertem Uran spielt die Schnellspaltung des $^{238}$U eine Rolle. Da bei der
Spaltung von $^{238}$U-Kernen gar etwa $1.5\%$ der Neutronen verz\"ogert freigesetzt werden, kann
das effektive $\beta$ in Versuchsreaktoren wie zum Beispiel dem CROCUS der Eidgen\"ossischen Technischen
Hochschule in Lausanne, welcher nur schwach angereichertes Uran enth\"alt, einen Wert von $0.78\%$ erreichen.
Dabei spielt auch der Umstand eine Rolle, dass prompte Neutronen auf Grund ihrer hohen Energie
(E$_{kin}\simeq 2$MeV) mit einer etwas
gr\"osseren Wahrscheinlichkeit aus einem kleinen Versuchsreaktor herausgestreut werden als die etwas
energie\"armeren verz\"ogerten Neutronen (E$_{kin}\simeq 0.5$MeV). Da in kommerziellen Reaktoren im Verlaufe
ihres Betriebs Plutonium mit $\beta_{Pu-239}=0.21\%$ erbr\"utet wird, nimmt der Anteil verz\"ogerter Neutronen
im Verlaufe des Leistungsbetriebs wiederum leicht ab. Eine Abnahme von $\beta_{BOC}=0.6\%$ auf
$\beta_{EOC}=0.5\%$ (BOC/EOC: beginning/end of cycle) im Verlaufe eines Betriebsjahres
w\"are f\"ur einen Gigawatt-Leichtwasserreaktor mit einem Durchschnittskern, welcher Brennelemente
unterschiedlichen Alters enth\"alt, recht typisch.
\\

\noindent Die obige Form der kinetischen Grundgleichungen vernachl\"assigt offensichtlich den Einfluss einer
Neutronenquelle, die unabh\"angig
von der Kernspaltung mit zeitlich etwa konstanter Rate $Q_{th}$ thermische Neutronen im Reaktor erzeugt.
Diese N\"aherung ist
im einem Leistungsreaktor (mit $\rho >0$ im Anfahrbereich oder $\rho=0$ im station\"aren Leistungsbereich)
meistens sinnvoll, wenn die Neutronenproduktionsrate durch Kettenreaktion die Neutronenerzeugungsrate durch
Zerf\"alle um mehrere Zehnerpotenzen \"ubersteigt.  Im Anfahrbereich ist die Leistung des Reaktors noch so gering,
dass diese nicht auf den Wert der Reaktivit\"at merklich r\"uckkoppelt.
\\

\noindent  F\"ur $\rho \le 0$ muss Gleichung (\ref{gl1}) ersetzt werden durch
\begin{equation}
\dot{n}  =  \frac{\rho - \beta}{\Lambda} n  \, \, +  \lambda C  + Q_{th} \, , \label{gl1_Q}
\end{equation}
mit entsprechenden Konsequenzen, die sp\"ater diskutiert werden.
\\

\noindent Im station\"aren Zustand gilt nun $\dot{n} = \dot{C} = 0$, \emph{per definitionem} sind in einem
station\"aren Zustand physikalisch messbare Gr\"ossen zeitlich unver\"anderlich.
Gem\"ass Gleichung (\ref{verhaeltnis}) gilt im station\"aren Zustand 
$C = \frac{\beta}{\lambda \Lambda} n$.
Befindet sich ein Reaktor also im station\"aren Zustand und wird in diesem gest\"ort, so gilt anf\"anglich
\begin{equation}
 \frac{\beta}{\Lambda} n    =   \lambda C \, .
\end{equation}
Ist weiter die Reaktivit\"at $\rho$ von \"ahnlicher Gr\"ossenordnung wie $\beta$, so ist auch
$\frac{\beta}{\Lambda} n$ von \"ahnlicher Gr\"os\-sen\-ord\-nung wie $ \frac{\rho - \beta}{\Lambda} n$.
Daher \"andert sich die Neutronenzahl zeitlich um \"ahnliche Betr\"age wie die der Vorl\"au\-fer (solange
der Reaktor \emph{prompt unterkritisch} bleibt: $\rho < \beta$ beziehungsweise $\rho_p = \rho - \beta < 0$),
doch diese sind etwa tausendfach zahlreicher. Entsprechend reagiert die Vor\-l\"au\-fer\-zahl etwa tausendfach
langsamer als die Neutronenzahl auf St\"orungen des station\"aren Zustandes. Diese Aussage gilt allerdings
nicht mehr im \emph{prompt \"uberkritischen} Reaktor mit $\rho > \beta$, in dem die Neutronenzahl schon kurze
Zeit nach dem Erreichen der gef\"ahrlich hohen Reaktivit\"at die Vorl\"auferzahl \"ubertreffen kann.
\\

\noindent Die entsprechenden Zeitskalen im prompt unterkritischen Reaktor sind f\"ur den Be\-reich $0 \leq \rho < 0.9 \beta$
gegeben durch die \emph{prompte Zeitkonstante} $\omega_{pr}$\\
%($\rightarrow$ T/RK 2-4-3)
\begin{equation}
t_{e,pr} = \frac{1}{\omega_{pr}}=\frac{\Lambda}{\beta-\rho}  \simeq \frac{10^{-4} \, \mbox{s}}{0.01 \ldots 0.001}
=0.01 \ldots 0.1 \, \mbox{s} \, ,
\end{equation}
dabei nennt man $t_{e,pr}$ auch die \emph{prompte Zeitverz\"ogerung},
die sogenannte verz\"ogerte Zeitkonstante ist gegeben durch $\omega_{v}=1/t_{e,v}=\lambda \simeq 0.1$ s$^{-1}$.
$t_{e,v}$ heisst verz\"ogerte Zeitverz\"ogerung, auf dieser Zeitskala reagieren die Vorl\"aufer in der
Eingruppenn\"aherung auf \"Anderungen der Neutronenbilanz.
\\

\noindent Bevor die exakten L\"osungen der Eingruppengleichungen untersucht werden, soll
an dieser Stelle kurz die Situa\-tion diskutiert werden, bei der die Reaktivit\"at $\rho_0$ in einem
Reaktor mit $n_0$ Neutronen abrupt einen neuen Wert $\rho$ annimmt. Vereinfachend soll dabei gelten, dass innerhalb des
betrachteten Zeitrahmens die Vorl\"auferzahl als eine Konstante betrachtet werden darf. Aus
\begin{equation}
\dot{n}  =  \frac{\rho - \beta}{\Lambda} n  \, \, +  \lambda C =   \frac{\rho_p}{\Lambda} n  \, \, +  \lambda C \label{sdgl}
\end{equation}
mit der sogenannten \emph{prompten Reaktivit\"at} $\rho_p <0$ folgt so
\begin{equation}
n(t)=\biggl( n_0+\frac{\lambda  \Lambda}{\rho_p} \cdot C \biggr) e^{\frac{\rho_p}{\Lambda}t} -
\frac{\lambda \Lambda}{\rho_p} \cdot C \, .
\end{equation}
Dieser zeitliche Verlauf der Neutronenzahl l\"ost offensichtlich Gleichung (\ref{sdgl}) mit der Anfangswertbedingung
$n(t=0)=n_0$. Die L\"osung zerf\"allt mit der charakteristischen Zeitskala $t_{e,pr}$ gegen den (quasi)station\"aren
Gleichgewichtswert $n(t \gg t_{e,pr})= (\lambda \Lambda / |\rho_p|) \cdot C$. Letztlich reagiert aber auf einer
gr\"osseren Zeitskala auch die Vorl\"auferzahl auf die ver\"anderten Bedingungen.
\\

\noindent Bei einer schnellen \"Anderung der Reaktivit\"at eines \emph{prompt unterkritischen} Reaktors
schwingt die Neutronenzahl also innerhalb von Bruchteilen einer Sekunde wieder in ein \emph{quasistation\"ares
Gleichgewicht} mit den Vorl\"aufern ein, sodass n\"aherungsweise gilt
\begin{equation}
n=\frac{\lambda \Lambda}{|\rho_p|} \cdot C = \frac{\lambda \Lambda}{\beta-\rho} \cdot C \, .
\end{equation}
Mit  gr\"osserer Verz\"ogerung stellt sich schliesslich auch eine \"Anderung der Vorl\"auferzahl ein.
N\"ahert sich der Wert der Reaktivit\"at dem prompt kritischen Wert $\rho=\beta$, so werden die
oben gemachten Annahmen unzul\"assig. Die prompte Reaktivit\"at $\rho_p=\rho-\beta$
misst sozusagen den Abstand von dieser Schwelle.

%------------------------------------------------------------------------------------------------------------------------------------

\subsubsection*{L\"osung (mit spezieller Betrachtung des verz\"ogert \"uberkritischen Zustands $0< \rho < \beta$)}

Um die vereinfachten kinetischen Grundgleichungen \emph{exakt} zu l\"osen, leiten wir Gleichung (\ref{gl1})
zuerst noch einmal ab und erhalten flugs
\begin{equation}
\ddot{n} =  \frac{\rho - \beta}{\Lambda} \dot{n} + \lambda \dot C \label{komp}
\end{equation}

\noindent Aus der Summe von Gleichung (\ref{gl1}) und (\ref{gl2}) folgt aber sofort
\begin{equation}
\dot{C}+\dot{n} =  \frac{\rho }{\Lambda} n  \quad \mbox{oder} \quad \lambda \dot C = \frac{\lambda \rho}{\Lambda} n
- \lambda \dot n \, , \label{short}
\end{equation}
ersetzen wir also $\lambda \dot C$ in Gleichung (\ref{komp}) durch den Ausdruck aus Gleichung (\ref{short}),
so ergibt sich
\begin{equation}
\ddot n + \frac{\beta - \rho}{\Lambda} \dot n + \lambda \dot n - \frac{\lambda \rho}{\Lambda} n = 0 \, . \label{finale}
\end{equation}
Diese Gleichung l\"asst sich durch einen Exponentialansatz einfach l\"osen - wir setzen $n(t)=n_0 \cdot e^{\omega \cdot t}$
und erhalten wegen $\frac{d}{dt} e^{\omega \cdot t}=\omega \cdot e^{\omega \cdot t}$
\begin{equation}
\ddot{n}= \omega^2 \cdot n =\omega^2 \cdot n_0 \cdot e^{\omega \cdot t} \quad \mbox{und} \quad
 \dot{n}= \omega \cdot n = \omega \cdot n_0 \cdot e^{\omega \cdot t} \, .
\end{equation}
Dies eingesetzt in Gleichung (\ref{finale}) ergibt
\begin{equation}
\omega^2 \cdot n_0 \cdot e^{\omega \cdot t} + \omega \, \Biggr( \frac{\beta - \rho}{\Lambda}  + \lambda  \Biggl)
\, n_0 \cdot e^{\omega \cdot t} - \frac{\lambda \rho}{\Lambda} \cdot n_0 \cdot e^{\omega \cdot t} =0 \, .
\end{equation}
Nat\"urlich k\"onnen die Faktoren $n_0$ und $e^{\omega \cdot t}$ aus dieser Gleichung weggek\"urzt werden.
Somit erhalten wir die sogenannte S\"akulargleichung f\"ur die \emph{reziproken Reaktorperioden}
\begin{equation}
\omega^2 +  \Biggr( \frac{\beta - \rho}{\Lambda}  + \lambda  \Biggl) \omega - \frac{\lambda \rho}{\Lambda} =0 \, . \label{sak}
\end{equation}
Aus der quadratischen Gleichung (\ref{sak})  ergeben sich n\"amlich zwei m\"ogliche L\"osungen f\"ur $\omega$, n\"amlich
\begin{equation}
\omega_{1,2}= \frac{1}{2} \Biggr( \frac{\rho-\beta}{\Lambda} - \lambda \Biggl) \pm
\frac{1}{2} \sqrt{\Biggr( \frac{\rho-\beta}{\Lambda} - \lambda \Biggl)^2+ 4 \frac{\lambda \rho}{\Lambda}} \, . \label{qua}
\end{equation}
Folgende Eigenschaft dieses Audruck ist bemerkenswert, {sofern}
die Reaktivit\"at des Reaktors positiv ist - diesen Fall wollen wir in der Folge zuerst diskutieren.
Der Wurzelterm ist n\"amlich betragsm\"assig
gr\"osser als der linke Klammerterm in Gleichung (\ref{qua}), da unter der Wurzel zum positiven Quadrat des
Klammerterms auch noch eine positive Gr\"osse $4 \lambda \rho / \Lambda$ hinzu addiert wird.
Daher ist $\omega_1$ (definiert mit dem Pluszeichen in der L\"osungsformel) immer positiv, und $\omega_2$ ("-")
immer negativ. Die beiden L\"osungskonstanten entsprechen also einer exponentiell ansteigenden und einer
exponentiell abfallenden Komponente in der Neutronenzahl, die sich im Reaktor \"uberlagern. Die exponentiell
ged\"ampfte L\"osung kann aber bei der Existenz einer positiven, sogenannten
\emph{stabilen Reaktorperiode} bald vernachl\"assigt werden, wenn also im \"uberkritischen
Reaktor die Neutronenzahl exponentiell zu wachsen beginnt. Offensichtlich ist $\omega_2$ mit der prompten
Zeitverz\"ogerung eng verkn\"upft; sie wird auch als \emph{transiente (reziproke) Reaktorperiode} bezeichnet.

\noindent Die allgemeinste L\"osung der Gleichung (\ref{finale}) lautet also 
\begin{equation}
n(t)=n_{ex} \cdot e^{\omega_1 \cdot t} + \tilde{n} \cdot e^{\omega_2 \cdot t} \,
\end{equation}
wobei der $e^{\omega_2 \cdot t}$-Anteil exponentiell zerf\"allt. Dieser beschreibt das quasistation\"are Einschwingen
der Neutronenzahl auf die Vorl\"auferzahl im Rahmen der Eingruppenn\"aherung.
\\

\noindent Weiter ist im Bereich $0 < \rho < 0.9 \cdot \beta$ (oder, ausgedr\"uckt durch die \emph{normierte Reaktivit\"at}
$\rho_n = \rho/\beta$, im Bereich $0<\rho_n < 90 \, \mbox{c} \!\!\! / \, = 0,9$ \$ )
der betragsm\"assige Wert von $\frac{\rho-\beta}{\Lambda}$
bedeutend gr\"osser als $\lambda$. Man kann daher vereinfachen:
\begin{equation}
\frac{1}{2} \sqrt{} = \frac{1}{2} \sqrt{\Biggr( \frac{\rho-\beta}{\Lambda} - \lambda \Biggl)^2+
4 \frac{\lambda \rho}{\Lambda}}
\simeq 
\frac{1}{2} \Biggl( \frac{\beta-\rho}{\Lambda} + \lambda \Biggr)
 \sqrt{1+\frac{4 \lambda \rho}{\Lambda} \frac{\Lambda^2}{(\beta-\rho)^2}} \, .
\end{equation}
Da $\frac{4 \lambda \rho \Lambda}{(\beta-\rho)^2}$ viel kleiner als eins ist und $\sqrt{1+x} \simeq 1 + \frac{1}{2} x$
f\"ur kleine $x$, folgt schliesslich\\
(wieder mit $\lambda \ll (\beta-\rho)/\Lambda)$
\begin{equation}
\frac{1}{2} \sqrt{} \simeq \frac{1}{2} \Biggl( \frac{\beta-\rho}{\Lambda} + \lambda \Biggr)
 \Biggr( 1 + \frac{2 \lambda \rho \Lambda}{(\beta-\rho)^2}
\Biggr) \, .
\end{equation}
Somit ist $\omega=\omega_1$ n\"aherungsweise durch die Formel
\begin{equation}
\omega =  \frac{1}{2} \Biggr( \frac{\rho-\beta}{\Lambda} - \lambda \Biggl) + 
\frac{1}{2} \Biggr( \frac{\beta-\rho}{\Lambda} + \lambda \Biggl) + \frac{\lambda \rho}{\beta - \rho} = \frac{\lambda \rho}
{\beta-\rho} \label{naefo}
\end{equation}
gegeben. %($\rightarrow$ T/RK 2-4-5 unten)
\\

\noindent Die stabile Reaktorperiode ist also im Rahmen der vereinfachten kinetischen Grundgleichungen n\"aherungsweise
gegeben durch
\begin{equation}
t_e = \frac{\beta-\rho}{\lambda \rho} \, , \label{naeh}
\end{equation}
die exakte Formel ist aber gem\"ass Gleichung (\ref{qua})
\begin{equation}
\omega= \frac{1}{2} \Biggr( \frac{\rho-\beta}{\Lambda} - \lambda \Biggl) +
\frac{1}{2} \sqrt{\Biggr( \frac{\rho-\beta}{\Lambda} - \lambda \Biggl)^2+ 4 \frac{\lambda \rho}{\Lambda}} \, .
\label{quaqua}
\end{equation}

\noindent Die stabile Reaktorperiode wird gleichwertig zur \emph{reziproken} stabilen Reaktorperiode $\omega$
\begin{equation}
t_e = \frac{1}{\omega}
\end{equation}
verwendet und beschreibt also die Zeit, in welcher sich die Neutronenzahl im eingeschwungenen anfahrenden
Reaktor ver-e-facht,
die \emph{Verdopp(e)lungszeit}, d.h. die Zeit, nach der sich im eingeschwungenen Zustand die
Neutronenzahl jeweils verdoppelt, ist entsprechend etwas kleiner
\begin{equation}
t_2=\ln(2) \cdot t_e = \frac{\ln(2)}{\omega} \simeq 0.69 \cdot t_e \, .
\end{equation}
Figur 1 stellt die Beziehung aus Gleichung (\ref{quaqua}) graphisch dar. Im angezeigten Bereich lassen sich
die Unterschiede zwischen Gleichung (\ref{naeh}) und (\ref{quaqua}) von Auge kaum ausmachen.
\begin{center}
\includegraphics[width=9.0cm]{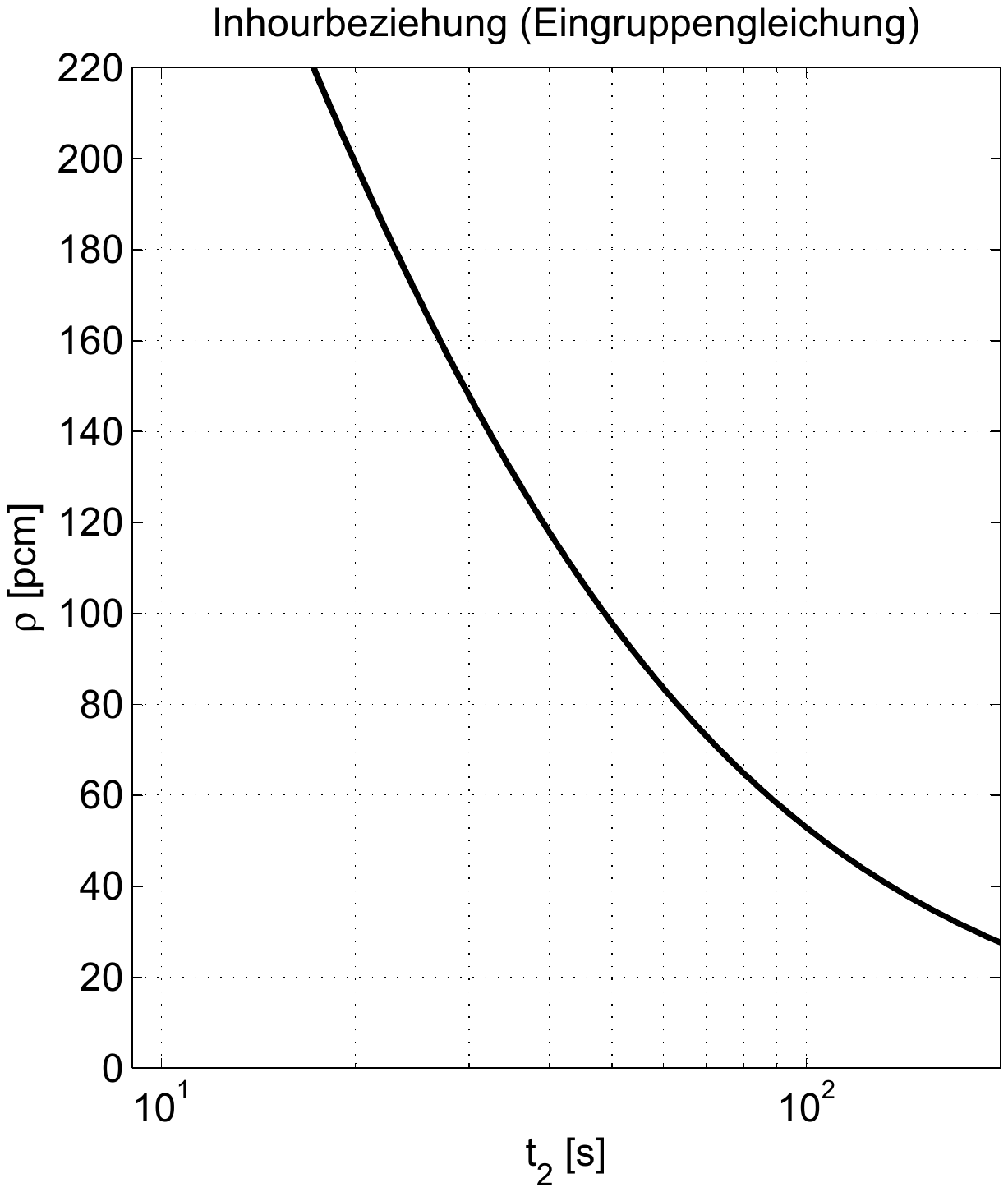}\\
\vspace{-1.0cm} Figur 1: Verdoppelungszeit versus Reaktivit\"at in der Eingruppenn\"aherung.
\end{center}

%------------------------------------------------------------------------------------------------------------------------------------

\subsubsection*{Inhourbeziehung}
Betrachten wir die sogenannte \emph{Inhourkurve} (Figur 2), welche den Zusammenhang zwischen Reaktivit\"at und
Verdoppelungszeit eines typischen Reaktors mit $\beta=0.64\%$ darstellt, so f\"allt prim\"ar eine gewisse Diskrepanz
im Bereich kleinerer Verdoppelungszeiten gegen\"uber Figur 1 auf.
Der Grund daf\"ur liegt in der simplen Tatsache begr\"undet, dass eine pr\"azisere Berechnung der Verdoppelungszeit
die Ber\"ucksichtigung aller 6 Vorl\"aufergruppen mit ihren individuellen Zerfallkonstanten $\lambda_i$ und
Produktionsanteilen $\beta_i$  voraussetzt.

Die vollen kinetischen Grundgleichungen
\begin{equation}
\dot{n}  =  \frac{\rho - \beta}{\Lambda} n  \, \, +  \sum \limits_{i=1}^{6} \lambda_i C_i  \, , \label{gl1v}
\end{equation}
\begin{equation}
 \dot{C_i}  =  \frac{\beta_i}{\Lambda} n  -   \lambda_i C_i  \,  , \quad  \,
\beta = \sum \limits_{i=1}^{6} \beta_i ,  \label{gl2v}
\end{equation}
lassen sich tats\"achlich durch einen Exponentialansatz l\"osen. Es stellt sich heraus, dass sich die
Neutronenzahl als Summe von sieben verschiedenen Exponentialfunktionen
\begin{equation}
n(t)= n_{ex} \cdot e^{\omega_1 \cdot t} + \sum \limits_{j=2}^{7} \tilde{n}_j \cdot e^{\omega_j \cdot t}
\label{seven}
\end{equation}
schreiben l\"asst. Eine analoge Aussage gilt f\"ur die Vorl\"auferzahlen.
Dabei folgt aus einer detaillierteren mathematischen Analyse, dass im Falle $\rho > 0$
\begin{equation}
\omega = \omega_1 > 0 \, \quad \mbox{und} \, \, \,  \omega_j < 0 \, \,  \, \mbox{f\"ur} \, \, \, j=2, \ldots 7 \, .
\end{equation}
\begin{center}
\includegraphics[width=9.0cm]{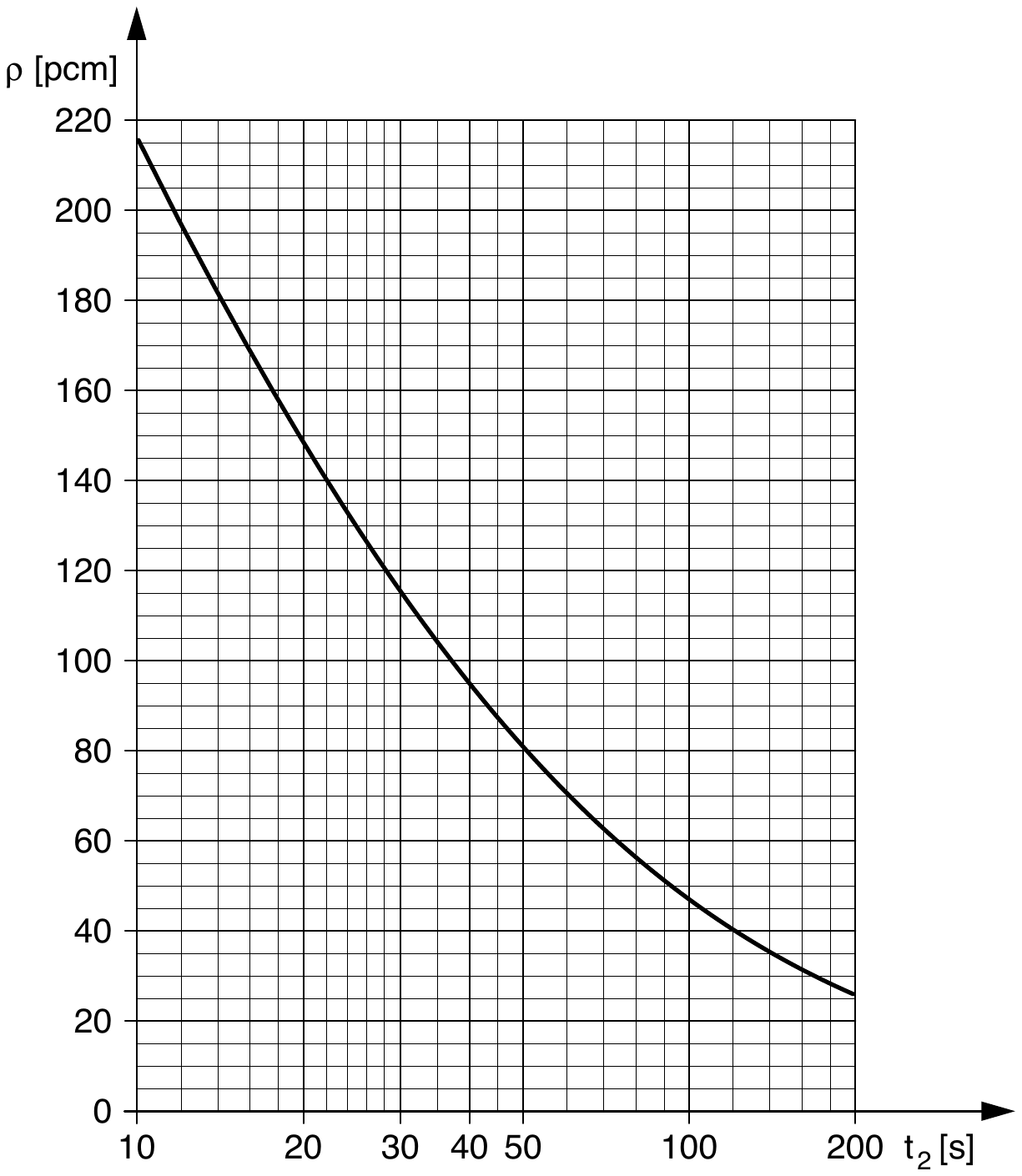}\\
\vspace{-1.0cm} Figur 2: Die Inhourkurve.
\end{center}

\noindent F\"ur $\rho<0$ sind alle $\omega_j$ negativ, im unterkritischen Reaktor zerf\"allt jede St\"orung
der Neutronenzahl, und so strebt diese jeweils gegen den durch die Neutronenquellen vorgegebenen Gleichgewichtswert
gem\"ass Gleichung (\ref{invn}).
F\"ur $\rho>0$ existiert aber eine stabile Reaktorperiode $\omega$, welche in der Inhourkurve (Figur 2)
abgelesen werden kann. Die restlichen Exponentialfunktionen fallen zeitlich stark ab und
beschreiben den Einschwingvorgang, der einem stabilen exponentiellen Anstieg der Neutronenzahl
zeitlich vorangeht. Nach dem Einschwingvorgang wird die Entwicklung der Neutronenzahl faktisch nur noch durch den
Term
\begin{equation}
n(t)=n_{ex} \cdot e^{\omega \cdot t}
\end{equation}
beschrieben. Offensichtlich ist $n(0)=n_{ex} + \sum \limits_{j=2}^{7} \tilde{n}_j$ die Startzahl der Neutronen,
wobei nat\"urlich bei unseren Betrachtungen immer $\rho=$konst. f\"ur $t \ge 0$ gilt. In einem Leistungsreaktor
ist es typischerweise so, dass eine Reaktivit\"atserh\"ohung aus dem kritischen Zustand $\rho=0$ zu
$\rho>0$ eine Erhitzung des Brennstoffs (und des K\"uhlmittels oder Moderators) bewirkt, welche die
Reaktivit\"at des Reaktors wieder reduziert. Durch diese stabilisierende R\"uckkopplung geht der Reaktor bei erh\"ohter
Leistung letztlich wieder in den kritischen Zustand \"uber.
\\

\noindent Die Zeitkonstante $\omega_2$ im Eingruppenmodell
kann wegen $\lambda \ll (\beta-\rho)/\Lambda$ in Gleichung (\ref{qua})
durch
\begin{equation}
\omega_2 \simeq - \frac{\beta -\rho}{\Lambda}
\end{equation}
angen\"ahert werden. Es handelt sich bei $\omega_2$
(bis auf ein Vorzeichen) n\"amlich um die prompte Zeitkonstante $\omega_{pr}$.
Sie repr\"asentiert im Wesentlichen die zus\"atzlichen 6 Zeitkonstanten $\omega_2, \ldots , \omega_7$ des
vollen Modells mit 6 Vorl\"aufergruppen. Diese Vereinfachung wird in gewissen F\"allen einer realistischen
Beschreibung nicht mehr gerecht.\\

\noindent Man kann zeigen, dass die Reaktivit\"at des Reaktors mit jeder Zeitkonstanten $\omega_j$ f\"ur
$j=1,...,7$ die exakte mathematische Beziehung (\emph{Inhour-Beziehung}, siehe Anhang)
\begin{equation}
\rho = \omega_j \cdot \Biggl( \Lambda + \sum \limits_{i=1}^{6} \frac{\beta_i}{\omega_j + \lambda_i} \Biggr)
\end{equation}
erf\"ullt. Kann in dieser Gleichung $\Lambda$ vernachl\"assigt werden, so erhalten wir f\"ur die Eingruppengleichungen
\begin{equation}
\rho=\omega \frac{\beta}{\omega+\lambda} \quad \rightarrow \quad \omega= \frac{\rho}{\beta-\rho} \lambda \, ,
\end{equation}
das heisst, wir gelangen wieder zu Gleichung (\ref{naeh}), wobei die negative reziproke Periode in der gemachten
N\"aherung unter den Tisch f\"allt.
\\

\noindent Die Inhour-Beziehung l\"asst eine unmittelbare Berechnung der Reaktivit\"at aus der Reaktorperiode
zu. Die Umkehrung dieses Vorgangs ist allerdings nur numerisch m\"oglich.
\\

\noindent Die stabile Reaktorperiode ist also n\"aherungsweise durch
\begin{equation}
t_e = \frac{\beta - \rho}{\rho} \cdot \frac{1}{\lambda} =  \frac{\beta - \rho}{\rho} \cdot l
\end{equation}
gegeben, und die Verdoppelungszeit folgt aus der Halbwertszeit $t_{1/2}= \ln(2) \cdot l \simeq 8.97$s der Vorl\"aufer
\begin{equation}
t_2 = \frac{\beta - \rho}{\rho} \cdot t_{1/2} \quad \leftrightarrow \quad \rho = \beta \cdot \frac{t_{1/2}}{t_2+t_{1/2}} \quad .
 \label{inhour_1}
\end{equation}
Misst man somit die stabile Neutronenzunahme in einem Reaktor durch Messung eines Neutronenflusses 
$\sim \Phi \propto n$, so l\"asst sich also mit Hilfe von Formel (\ref{inhour_1}) sofort die Reaktivit\"at n\"aherungsweise
berechnen. Genauere Aussagen gewinnt man aus der Inhourkurve, welche alle Vorl\"aufer ber\"ucksichtigt.\\

\noindent Beispiel: F\"ur eine Reaktivit\"at von 80pcm ergibt sich
\begin{equation}
t_{2} = \frac{0.0064-0.0008}{0.0008} \cdot 9\rm{s} = 63\rm{s} \,  ,
\end{equation}
in passabler \"Ubereinstimmung mit $t_2=50$s in der Inhourkurve.
\\

\noindent Der Ausdruck $Inhour$ geht nicht aus einen Forscher gleichen Namens zur\"uck, sondern
beruht historisch und vereinfacht gesagt auf der Tatsache, dass bei fr\"uhen Reaktorexperimenten ($\rightarrow$
Chicago Pile 1, kurz CP-1, der erste von Menschenhand gebaute Kernreaktor) der
Neutronenzuwachs  \emph{pro Stunde}, also in Beziehung zu \emph{inverse hours}, betrachtet wurde.
Entsprechend wird zur Beschreibung der Reaktivit\"at eines Reaktors auch die Einheit \emph{InHour} verwendet.
Dabei beschreibt 1ih oder 1inhr diejenige Reaktivit\"at eines Reaktors, bei welcher sich die Neutronenzahl
innerhalb einer Stunde ver-e-facht (etwa 2.3pcm).
\\

\noindent Da sich die Neutronenzahl im Bereich $0 \le \rho < \beta$ in zeitlich in kontrollierbarer Weise vergr\"ossert,
nennt man den entsprechenden Reaktivit\"atsbereich \emph{verz\"ogert \"uberkritisch.} Es ist also der Existenz der
Vorl\"aufer zu verdanken, dass sich die Leistung eines Reaktors zeitlich im Minutenbereich regeln l\"asst, sofern in einem
sicheren Reaktivit\"atsbereich unterhalb von $\sim 200$pcm operiert wird.

\subsection*{Allgemeine Diskussion der L\"osungen der kinetischen Grundgleichungen}
\subsubsection*{Reaktivit\"at als Ordnungsparameter}
In der Folge sollen nach der eingehenden Diskussion des \"uberkritischen Reaktors nun auch
noch die weiteren durch die Reaktivit\"at charakterisierbaren Reaktorzust\"ande diskutiert werden.
Ein Reaktor mit negativer Reaktivit\"at $\rho<0$ heisst also \emph{unterkritisch}, bei $\rho=0$ ist er \emph{kritisch};
gilt $\rho > 0$, so ist der Reaktor \emph{\"uberkritisch}.  Bei \"uberkritischen Reaktoren unterscheidet man
den kontrollierbaren Fall des Reaktors im Anfahrbereich mit $0<\rho <\beta$, welcher auch als \emph{verz\"ogert
\"uberkritisch} bezeichnet wird. Der prompt kritische ($\rho=\beta$) oder prompt \"uberkritische Zustand
($\rho > \beta$) muss auf jeden Fall vermieden werden. Eine Ausnahme bilden hier allerdings die sogenannten
TRIGA-Forschungsreaktoren (\emph{Training, Research, Isotopes, General Atomic}), welche in kontrollierter
Weise f\"ur einige Millisekunden in den prompt \"uberkritischen Zustand gebracht werden k\"onnen.

\subsubsection*{Quellbereich ($\rho < 0$)}
Im unterkritischen Bereich sind die beiden inversen Reaktorperioden
\begin{equation}
\omega_{1,2}= \frac{1}{2} \Biggr( \frac{\rho-\beta}{\Lambda} - \lambda \Biggl) \pm
\frac{1}{2} \sqrt{\Biggr( \frac{\rho-\beta}{\Lambda} - \lambda \Biggl)^2+ 4 \frac{\lambda \rho}{\Lambda}}
\end{equation}
negativ. Ohne Neutronenquelle w\"urde also die Neutronenzahl in einem Reaktor gegen Null zerfallen.
Be\-r\"uck\-sich\-tigt man eine potenziell vorhandene Neutronenquelle, so kann aus der allgemeinsten L\"osung der vereinfachten kinetischen
Grund\-glei\-chungen \emph{ohne} Quelle
\begin{equation}
n(t)=n_1 \cdot e^{\omega_1 \cdot t}+n_2 \cdot e^{\omega_2 \cdot t} 
\end{equation}
sofort eine L\"osung des Gleichungssystems
\begin{equation}
\dot{n}  =  \frac{\rho - \beta}{\Lambda} n  \, \, +  \lambda C  + Q_{th} \, \,  ,  \label{unter1}
\end{equation}
\begin{equation}
\dot{C}  =  \frac{\beta}{\Lambda} n -   \lambda C     \label{unter2}
\end{equation}
angegeben werden:
\begin{equation}
n(t)= - \frac{\Lambda \cdot Q_{th}}{\rho} + n_1 \cdot e^{\omega_1 \cdot t}+n_2 \cdot e^{\omega_2 \cdot t}  \quad ,
\label{partikulaer}
\end{equation}
\begin{equation}
C(t)=- \frac{\beta}{\lambda \rho} \cdot Q_{th} +
\Biggl( \frac{\omega_1}{\lambda}+ \frac{\beta-\rho}{\lambda  \Lambda} \Biggr) \cdot n_1 \cdot e^{\omega_1 \cdot t} +
\Biggl( \frac{\omega_2}{\lambda}+ \frac{\beta-\rho}{\lambda  \Lambda} \Biggr) \cdot n_2 \cdot e^{\omega_2 \cdot t}
\quad .
\end{equation}
F\"ur grosse Zeiten strebt diese L\"osung den station\"aren Werten
\begin{equation}
n_{stat} = - \frac{\Lambda \cdot Q_{th}}{\rho} \, , \quad 
C_{stat} = - \frac{\beta}{\lambda \rho} \cdot Q_{th} = \frac{\beta}{\lambda \Lambda} n_{stat}
\end{equation}
zu, welche sich sofort aus den station\"aren, zeitunabh\"angigen Bilanzgleichungen
\begin{equation}
\dot{n}=0=  \frac{\rho - \beta}{\Lambda} n  \, \, +  \lambda C  + Q_{th} \, , \quad 
\dot{C}=0= \frac{\beta}{\Lambda} n    -   \lambda C 
\end{equation}
ergeben. Offensichtlich ist die station\"are Neutronenzahl und die Vorl\"auferzahl im Reaktor proportional zu $-\frac{1}{\rho}$.
Entsprechend dauert es bei Reaktivit\"ats\"anderungen im unterkritischen Reaktor umso l\"anger, bis sich der Reaktor wieder
auf einen station\"aren Zustand eingeschwungen hat, je n\"aher man
sich an der kritischen Schwelle $\rho=0$ befindet. Wird die negative Reaktivit\"at schrittweise zu betragsm\"assig kleineren Werten
reduziert, so stellt sich nach einer Einschwingzeit eine schrittweise zunehmend gr\"osser werdende station\"are Neutronen- und Vorl\"auferzahl ein.
Der Aufbau dieser zunehmend gr\"osser werdenden Zahlen dauert dabei zunehmend l\"anger, da die eigentliche Kettenreaktion
im unterkritischen Reaktor immer defizit\"ar ist - das gekoppelte System der Neutronen und Vorl\"aufer wird
letztlich aus der Reaktorquelle erzeugt, die mit konstanter Rate Neutronen nachliefert.
Bildet man n\"amlich die Summe der Gleichungen (\ref{unter1}) und (\ref{unter2}), so ergibt sich
\begin{equation}
\dot{n}+\dot{C} = \frac{\rho}{\Lambda} n + Q_{th} \, . \label{summa}
\end{equation}
Der Term $\frac{\rho}{\Lambda} n$ ist aber negativ, da $\rho$ negativ ist. Die Gesamtzahl $\dot{n}+\dot{C}$ der Neutronen und
Vorl\"aufer zusammen, die f\"ur schwindendes $\rho$ immer gr\"ossere Werte annimmt, kann nur dank des
Quellterms $Q_{th}$ in Gleichung (\ref{summa}) wachsen.
\\

\noindent Bei einer Reaktivit\"ats\"anderung im Bereich betragsm\"assig kleiner Reaktivit\"aten ergibt sich
die Einschwingzeit $t_{es}$ eines Reaktors im Wesentlichen
aus dem am langsamsten zerfallenden Exponentialterm in Gleichung (\ref{partikulaer})
(siehe auch Glei\-chung (\ref{naefo})) mit reziproker Periode
\begin{equation}
0 > \omega_1= \frac{1}{2} \Biggr( \frac{\rho-\beta}{\Lambda} - \lambda \Biggl) +
\frac{1}{2} \sqrt{\Biggr( \frac{\rho-\beta}{\Lambda} - \lambda \Biggl)^2+ 4 \frac{\lambda \rho}{\Lambda}}
\simeq \frac{\lambda \rho}{\beta -\rho} \simeq \frac{\lambda \rho}{\beta} \quad (0<-\rho \ll \beta)
\end{equation}
zu
\begin{equation}
t_{es} =- \frac{1}{\omega_1} \simeq \frac{\beta}{\lambda  |\rho|} \, .
\end{equation}
F\"ur $\beta=0.0064=640$pcm und $1/\lambda=12.9$s ergibt sich daraus
$t_{es} \simeq \frac{137\rm{min}}{| \rho | \, \rm{in} \, \rm{pcm}} \, .$
Der tats\"achliche Einschwingvorgang dauert nat\"urlich (5-10mal) l\"anger als $t_{es}$, da die
Eingruppengleichung den Einfluss der lang\-lebigen Vor\-l\"au\-fer untersch\"atzt und da $t_{es}$ im
Wesentlichen die Zeit f\"ur eine Reduktion der Abweichung des Neutronen-Vorl\"aufer-Sys\-tems vom
Gleichgewichtszustand auf einen e-tel beschreibt.
\\

\noindent Wird die Reaktivit\"ats\"anderung schnell (prompt) durchgef\"uhrt, so vollzieht die Neutronenzahl
zuerst einen sogenannten prompten Sprung, bei welchem ein quasistation\"ares Gleichgewicht zwischen Neutronen
und Vorl\"aufer herbeigef\"uhrt wird. Dieser Vorgang wird mit der betragsm\"assig gr\"osseren Zeitkonstante
$\omega_2$ zeitlich beschrieben. Bei kleinen Reaktivit\"as\"anderung ist der prompte Sprung weniger ausgepr\"agt.
Quantitative Betrachtungen zum prompten Sprung finden sich im
Abschnitt \emph{Schnelle Reaktivit\"ats\"anderungen}.

\subsubsection*{Kritischer Reaktor: $\rho=0$}
Im kritischen Reaktor vereinfachen sich die kinetischen Grundgleichungen mit $\rho=0$ zu
\begin{equation}
\dot{n}  =  -\frac{\beta}{\Lambda} n  \, \, +  \lambda C +Q_{th} \, , \quad
\dot{C}  =  \frac{\beta}{\Lambda} n   -   \lambda C  \quad , \label{gl2k}
\end{equation}
und bildet man die Summe dieser beiden Gleichungen, so folgt
\begin{equation}
\dot{C}+\dot{n}= Q_{th} \, ,
\end{equation}
also
\begin{equation}
C+n= N_0 + Q_{th} \cdot t \, .
\end{equation}
Da die Gesamtzahl von Vorl\"aufern und Neutronen offensichtlich linear mit der Zeit ansteigt, da die
Kettenreaktion eigentlich selbsterhaltend ist, die Neutronenquelle aber immer mehr Neutronen in den
Reaktor einspeist, liegt es nahe, sowohl f\"ur die Anzahl der Vorl\"aufer sowie f\"ur die Anzahl der Neutronen
einen zeitlich linear ansteigenden Verlauf anzusetzen:
\begin{equation}
n(t)=n_0 + Q_1 \cdot t \, , \quad C(t)=C_0 + Q_2 \cdot t \, , \quad Q_1+Q_2=Q_{th} \, .
\end{equation}
Eingesetzt in die Neutronenbilanzgleichung (\ref{gl2k}) liefert der Ansatz
\begin{equation}
Q_1 = -\frac{\beta}{\lambda} n_0 - \frac{\beta}{\lambda} Q_1 \cdot t + \lambda C_0 + \lambda Q_2 \cdot t + Q_{th} \, .
\end{equation}
Die in $t$ linearen Terme m\"ussen sich offensichtlich aufheben, also gilt
\begin{equation}
\frac{\beta}{\Lambda} Q_1 = \lambda Q_2 \quad \rightarrow \quad Q_2= \frac{\beta}{\lambda \Lambda} Q_1 = Q_{th}-Q_1 \, ,
\end{equation}
\begin{equation}
Q_1 =\frac{Q_{th}}{1+\frac{\beta}{\lambda \Lambda}} \, , \quad Q_2 = \frac{Q_{th}}{1+\frac{\lambda \Lambda}{\beta}} \, .
\end{equation}
In einem kritischen Reaktor \emph{mit} Quelle nimmt die Neutronenzahl nach dem Einschwingen linear zu, und zwar in einem
Zahlenverh\"altnis zu den Vorl\"aufern, welches bereits vom station\"aren, unterkritischen Fall her bekannt ist.
Die Anfangswerte $n_0$ und $C_0$ k\"onnen geeignet gew\"ahlt werden, sodass die kinetischen Grundgleichungen
erf\"ullt werden.
\\

\noindent Man beachte, dass sich zur linearen L\"osung der kinetischen Grundgleichungen
eine exponentiell zer\-fallende L\"osung $\sim e^{-\omega_{krit} \cdot t}$  mit
inverser Periode $\omega_{krit}= \beta/\Lambda+\lambda \approx \beta/\Lambda$ hinzuaddieren l\"asst.
Der Wert dieser Frequenz folgt wiederum aus
Gleichung (\ref{qua}) (mit entsprechendem Minuszeichen) und ist auch f\"ur den sogenannten inhomogenen Fall
der kinetischen Grundgleichungen mit Neutronenquelle g\"ultig. Der zerfallende L\"osungsanteil beschreibt
den prompten Einschwingvorgang der Neutronen unmittelbar nachdem der Reaktor kritisch wurde.

\subsubsection*{Prompt kritischer Reaktor ($\rho=\beta$)}
Die inversen Reaktorperioden lassen sich durch
\begin{equation}
\omega_{1,2}=- \frac{\lambda}{2} \pm \sqrt{\frac{\lambda^2}{4}+\frac{\lambda \beta}{\Lambda}} \simeq
\pm \sqrt{\frac{\lambda \beta}{\Lambda}}
\end{equation}
approximieren. Die positive Reaktorperiode beschreibt einen starken Anstieg der Neutronenzahl auf der prompt kritischen
Zeitskala $t_{pk}=\sqrt{\frac{\Lambda}{\bar{\lambda} \beta}} \simeq 0.2$s. Da auf dieser kurzen Zeitskala die Rolle der langlebigen
Vorl\"aufer durch die mittlere Zerfallskonstante $\lambda \simeq 0.1$s$^{-1}$ \"ubergewichtet wird,
muss $t_{pk}$ mit einer effektiven Zerfallskonstanten $\bar{\lambda} \simeq 0.407$s$^{-1}$ berechnet werden.

\subsubsection*{Prompt \"uberkritischer Reaktor ($\rho > \beta$)}
Im prompt \"uberkritischen Reaktor kann der Einfluss der Vorl\"aufer eigentlich vernachl\"assigt werden, da die prompten Neutronen allein
sich \"ausserst schnell vervielfachen. Ist $\rho$ hinreichend gross, sodass gilt
\begin{equation}
\frac{\rho_p}{\Lambda} = \frac{\rho-\beta}{\Lambda} \gg \lambda \, , 
\end{equation}
so ist
\begin{equation}
\omega_1 \simeq \frac{1}{2} \frac{\rho_p}{\Lambda} + \frac{1}{2} \sqrt{\frac{\rho_p^2}{\Lambda^2}}=\frac{\rho_p}{\Lambda} \, .
\end{equation}
Bereits f\"ur $\rho=2\beta$ oder $\rho_p=\beta$ ist $1/\omega_1 \simeq 0.016$s, d.h. in dieser Zeit ver-e-facht
sich die Neutronenzahl im Reaktor. Dieses Resultat ergibt sich auch aus der Neutronenbilanzgleichung (\ref{gl1}), wenn der Vorl\"auferterm
$\lambda C$ vernachl\"assigt wird:
\begin{equation}
\dot{n}= \frac{\rho_p}{\Lambda} n \quad \rightarrow \quad n(t) = n_0 \cdot e^{\frac{\rho_p}{\Lambda} \cdot t} \quad .
\end{equation}

\noindent Nat\"urlich ist der prompt (\"uber)kritische Zustand eines Reaktors in jedem Falle zu
vermeiden, da die ent\-spre\-chen\-de Leistungsexkursion zur Kernschmelze f\"uhren kann.
Auf der obigen sehr kurzen Zeitskala ist es allerdings nicht unbedingt statthaft, von einer Gesamtreaktivit\"at
(\emph{lumped reactivity}) des Reaktors zu sprechen (siehe Abschnitt \emph{Diffusion}). Zudem sind moderne
Reaktoren so ausgelegt, dass R\"uckkopplungseffekte bei einem Reaktivit\"atsanstieg die Reaktivit\"at
rechtzeitig zu binden verm\"ogen, sodass Schlimmeres verhindert wird. Dieses inh\"arent sichere Verhalten
war in den RBMK-1000-Reaktoren ($\rightarrow Tschernobyl$) nicht richtig implementiert. In modernen Leistungsreaktoren
stellt vielmehr die Abf\"uhrung des Nachzerfallsw\"arme, welche durch radioaktive Zerf\"alle von Produkten
des Reaktorbetriebs auch nach Unterbrechung der Kettenreaktion entsteht, ein potentielles Sicherheitsproblem
dar ($\rightarrow$ \emph{Fukushima}).

%------------------------------------------------------------------------------------------------------------------------------------

\subsection*{Schnelle Reaktivit\"ats\"anderungen}
Bei einer schnellen Reaktivit\"ats\"anderung ($\rho_0=0$ f\"ur $t<0$ $\rightarrow$ $\rho_1=\rho$ f\"ur $t>0$)
in einem kritischen Reaktor (zum Beispiel durch Stabeinwurf)
wird die Neutronenzahl $n_0$ f\"ur $t<0$ abrupt zu $n_1$ f\"ur $t>0$ ver\"andert, w\"ahrend das tr\"age Reservoir
der Vorl\"aufer von der Reaktivit\"ats\"anderung
vorerst wenig beeinflusst wird. Innerhalb k\"urzester Zeit stellt sich wieder ein quasistation\"ares Gleichgewicht
zwischen den Neutronen und den Vorl\"aufern ein, und es gilt in guter N\"aherung wie im station\"aren Zustand
\begin{equation}
\frac{\beta-\rho_0}{\Lambda} n_0 = \frac{\beta}{\Lambda} n_0= \lambda C = \frac{\beta-\rho_1}{\Lambda}
n_1=  \frac{\beta-\rho}{\Lambda} n_1 \, ,
\end{equation}
also ist in guter N\"aherung
\begin{equation}
n_1= \frac{\beta}{\beta-\rho} n_0 \, . \label{sprung}
\end{equation}
Suggestiver kann man f\"ur den prompten Sprung mit den prompten Reaktivit\"aten auch schreiben
($\rho_p=\rho-\beta$)
\begin{equation}
\rho_{p,0} \cdot n_0 = \rho_{p,1} \cdot n_1 \, . 
\end{equation}
In Rahmen der Eingruppenn\"aherung wird der zeitliche Verlauf der Neutronenzahl durch ($\rho_0=0$)
\begin{equation}
n(t)=n_0 \Biggl[ \frac{\beta}{\beta-\rho} e^{\frac{\lambda \rho}{\beta-\rho}t} -
\frac{\rho}{\beta-\rho} e^{-\frac{\beta-\rho}{\Lambda}t} \Biggr] 
\end{equation}
beschrieben. Es gilt $n(t=0)=n_0$. Da der zweite Exponentialterm sehr schnell zerf\"allt - die Zeitkonstante
$\frac{\Lambda}{\beta - \rho}$ (die prompte Zeitverz\"ogerung) ist im verz\"ogert kritischen Bereich von der Gr\"ossenordnung von
Bruchteilen einer Sekunde - schwingt sich die Neutronenzahl sehr schnell auf
\begin{equation}
n(t) \simeq n_0  \frac{\beta}{\beta-\rho} e^{\frac{\lambda \rho}{\beta-\rho}t} \, .
\end{equation}
ein. Der erste Exponentialterm beschreibt das Zerfallen oder das kontrollierte Anwachsen der Neutronenzahl im verz\"ogert
kri\-ti\-schen Reaktor mit stabiler Reaktorperiode. Allerdings stellt sich bei realistischer Ber\"uck\-sichti\-gung
aller Vorl\"aufer die stabile Reaktorperiode erst nach einem Einschwingvorgang ein, dessen Zeitskala auch mit
den verschiedenen Lebensdauern der Vorl\"aufer verkn\"upft ist.
\\

\noindent Bestimmt man in einem anf\"anglich kritischen Reaktor durch Neutronenflussmessungen
die relative Abnahme der Neutronenpopulation bei einer prompten Reaktivit\"ats\"anderung, so kann
\"uber Gleichung (\ref{sprung}) auf die tat\-s\"ach\-lich erfolgte Reaktivit\"ats\"anderung
geschlossen werden. Dies erm\"oglicht die Eichung von Steuerst\"aben als reaktivit\"atssteuernde
Elemente.

\subsection*{Reaktorschnellabschaltung}
Befindet sich ein Reaktor vor einer Schnellabschaltung im kritischen station\"aren Zustand, so l\"asst sich die Anzahl der
Vorl\"aufer in der i-ten Vorl\"aufergruppe mit Hilfe der Vorl\"auferbilanzgleichung (\ref{gl2}) sofort angeben:
\begin{equation}
\dot{C}=0 \quad \rightarrow \quad C_i= \frac{\beta_i}{\lambda_i \cdot \Lambda} n  \, .
\end{equation}
Durch die starke Reaktivit\"atserniedrigung auf $\rho \ll 0$ f\"allt die urspr\"unglich vorhandene Neutronenzahl $n_0$  prompt
auf den mit den Vorl\"aufern im quasistation\"aren Gleichgewicht befindlichen Wert
\begin{equation}
n_1 = \frac{\beta}{\beta-\rho} n_0 \, . \label{vorweg}
\end{equation}
Anschliessend zerfallen die Vorl\"aufer mit verschiedenen Zerfallskonstanten und setzen noch Neutronen frei.
Deren Anzahl berechnet sich zu ($n(0)=n_1$)
\begin{equation}
n(t)=\frac{\beta}{\beta-\rho} n_0 \sum \limits_{i=0}^{6} \frac{\beta_i}{\beta} e^{-\lambda_i \cdot t} \, .
\label{Abschaltung}
\end{equation}
In der obigen Formel wird der geringe Effekt der erneuten Vorl\"aufererzeugung durch die Freisetzung von
Neutronen und nachfolgender Kernspaltung durch die noch vorhandenen Vorl\"aufer vernachl\"assigt.
In guter Analogie mit der obigen N\"aherung w\"urde man in Gleichung (\ref{seven}) entsprechend
$\omega_j=-\lambda_{7-j}$ setzen,
wobei $\omega_7 \simeq -(\beta-\rho)/\Lambda$ in Gleichung (\ref{Abschaltung}) fehlt und die
Zeitskala f\"ur den durch Gleichung (\ref{vorweg}) vorweggenommenen prompten Sprung beschreibt.
\begin{center}
\includegraphics[width=8.0cm,angle=-90]{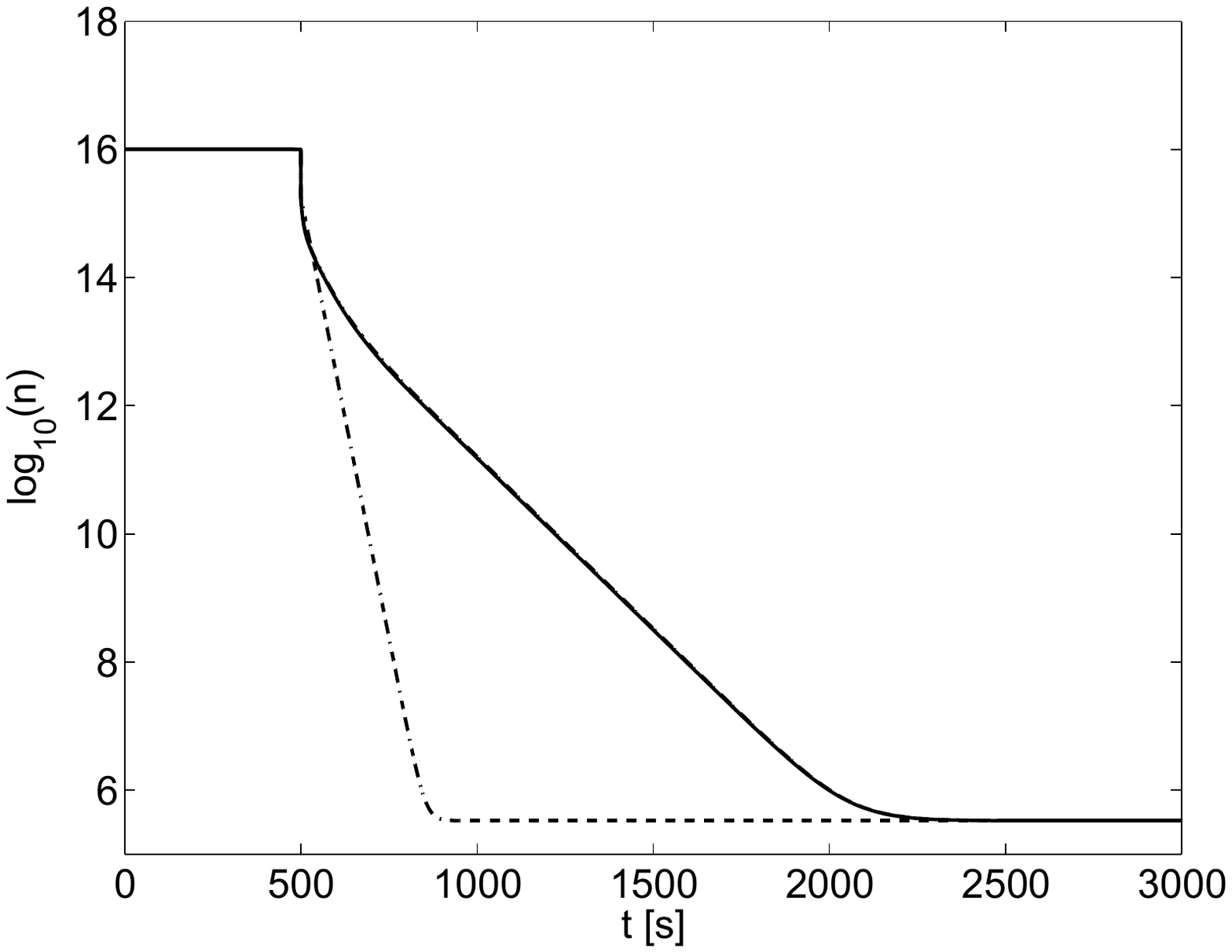}\\
\vskip 0.2cm Figur 3: Idealisierte schnelle Reaktorschnellabschaltung bei t=500s.
\end{center}

\noindent Der langlebigste Vorl\"aufer ist $^{87}$Br. Nachdem die kurzlebigeren Vorl\"aufer zerfallen sind, dominiert
$^{87}$Br die obige Summe und nach etwa einer Minute gilt (mit $b_6$ f\"ur $^{235}$U)
\begin{equation}
n(t) \simeq \frac{\rho}{\rho-\beta} n_0 \cdot b_6 \cdot e^{-\lambda_6 \cdot t} \quad \mbox{mit} \quad
b_6=\beta_6/\beta=0.032, \quad \lambda_6=1/(80.387\mbox{s}) \, .
\end{equation}
Dieser anf\"anglich exponentielle Zerfall der Neutronenzahl stabilisiert sich schliesslich nach einigen Minuten
bei dem endlichen Wert, der durch die Quellst\"arke der Reaktors vorgegeben wird.
\\

\noindent Figur 3 zeigt den numerisch berechneten Verlauf der Neutronenzahl nach einer schnellen
Reaktivit\"ats\-ver\-min\-derung um 3000pcm aus dem kritischen Zustand. Dabei ber\"ucksichtigt die ausgezogene
Linie die vollst\"andigen kinetischen Punktgleichungen mit sechs Vorl\"aufergruppen.
Die  Reaktivit\"at\-serniedri\-gung
$0\rm{pcm} \rightarrow -3\%$ induziert die prompte Reduktion der Neutronenzahl,
gefolgt vom neutronenfreisetzenden Zerfall der Vorl\"aufer. Wenige Minuten nach der Abschaltung tr\"agt nur noch
der $^{87}$Br-Zerfall merklich zur Neutronenzahl bei, schliess\-lich schwingt diese im Quellbereich auf
einen konstanten Wert ein. Die verwendeten Parameter betragen  $Q_{th}=10^8$s$^{-1}$, $\Lambda=10^{-4}$s,
$\beta=0.0064$ und $n_0=10^{16}$, die Vorl\"aufergruppenparameter nach Keepin \cite{Keepin} finden sich im
Anhang. Die gestrichelte Linie zeigt das Resultat im Rahmen der Eingruppenn\"aherung, welche lediglich zwei Reaktorperioden
beziehungsweise eine mittlere Vorl\"auferzerfallskonstante ber\"ucksichtigt. 

\subsection*{Prinzip des Reaktivit\"atsmeters}
Durch stetige \"Uberwachung des Neutronenflusses in einem Reaktor ist es prinzipiell
m\"oglich, aus dem zeitlichen Verlauf des Flusses, welcher im Rahmen der Punktmodells
proportional zur Neutronenzahl ist, die aktuelle Reaktivit\"at des Reaktors zu berechnen.\\

\noindent Die in der Folge als Beispiel diskutierte sogenannte Methode der inversen Reaktorkinetik,
deren mathematischen Grundlagen unten dargestellt sind,
l\"asst unter anderem die Bestimmung der Charakteristik eines Feinregelstabes in einer einzigen
Messung durch Einfahren des Steuerstabes mit konstanter Geschwindigkeit in den Reaktor zu.
Zwingende Vor\-aussetzung dazu ist nat\"urlich das Vorhandensein eines hinreichend starken Neutronensignals.
Zur Messung sehr grosser Reaktivit\"atsst\"orungen eignet sich die Methode letztlich nicht mehr,
da die im Punktmodell gemachten N\"aherungen dann unzul\"assig werden.
Immerhin kann die inverse Reaktorkinetik zum Aufsuchen eines guten Detektorplatzes
verwendet werden: An einem guten Ort wird die gemessene Reaktivit\"at nach einem
Reaktivit\"atssprung konstant verlaufen. Je schlechter die Messposition,
desto st\"arkere Zeitabh\"angigkeit (Schwingungen, sogenannte 'Harmonische')
zeigt der aus den Messungen berechnete Reaktivit\"atsverlauf nach einem abgeschlossenen
Reaktivit\"atssprung.\\

\noindent Ausgehend von den punktkinetischen Grundgleichungen f\"ur die (gewichteten) Neutronen- und
Vor\-l\"au\-fer\-zah\-len
\begin{equation}
\frac{dn}{dt} = \frac{\rho - \beta}{\Lambda} n(t) + \sum_{i=1}^{6} \lambda_i C_i (t) \, ,
\end{equation}
\begin{equation}
\frac{dC_i}{dt} = \frac{\beta_i}{\Lambda} n(t) - \lambda_i C_i (t) \, ,
\end{equation}
k\"onnen der Anteil $\beta$ verz\"ogerter Neutronen und die Generationszeit $\Lambda$ dazu verwendet werden,
diese Glei\-chungen zu reskalieren, und misst man also die (normierte) Reaktivit\"at $\rho$ in $\$ $, also in Einheiten von $\beta$,
durch die Definition $\rho_n= \rho^* = \rho/\beta$ und
$\Lambda^* = \Lambda/\beta$, so folgt 
\begin{equation}
\frac{dn}{dt} = \frac{\rho^*- 1}{\Lambda^*} n(t)  + \sum_{i=1}^{6} \lambda_i C_i (t) \, ,
\label{pker1}
\end{equation}
\begin{equation}
\frac{dC_i}{dt}= \frac{b_i}{\Lambda^*} n(t) - \lambda_i C_i (t) \, , \quad b_i=\frac{\beta_i}{\beta} \, ,
\label{pker2}
\end{equation}
und aus Gleichung (\ref{pker1}) erh\"alt man
\begin{equation}
\rho^* = 1 + \frac{\Lambda^*}{n(t)} \biggl\{ \frac{dn}{dt} -  \sum_{i=1}^{6} \lambda_i C_i (t) \biggr\} \, .
\end{equation}
Zur Zeit $t=0$ gilt im station\"aren Reaktor
\begin{equation}
\frac{dC_i}{dt} \biggr|_{t=0} = 0 \, \rightarrow \, C_i (0) = \frac{b_i}{\lambda_i \Lambda^*} n(0) \, ,
\end{equation}
und durch Integration der inhomogenen linearen Differenzialgleichung (\ref{pker2}) ergibt sich sofort
\begin{equation}
C_i(t) = C_i (0) e^{-\lambda_i t} + \frac{b_i}{\Lambda^*} \int_{0}^{t}
e^{-\lambda_i (t-t')} n(t') dt' \, .
\end{equation}\\

\noindent Um nun die Gleichungen (\ref{pker1}) und (\ref{pker2}) zu diskretisieren, berechnet man
\begin{displaymath}
C_i (t+\Delta t) = \frac{b_i}{\Lambda^*} \int_{0}^{t+ \Delta t} e^{-\lambda_i (t+\Delta t -t')} n(t') dt'
+C_i(0) e^{-\lambda_i (t+\Delta t)}
\end{displaymath}
\begin{displaymath}
= \frac{b_i}{\Lambda^*} e^{-\lambda_i \Delta t}  \int_{0}^{t} e^{-\lambda_i (t -t')} n(t') dt'
+ \frac{b_i}{\Lambda^*} e^{-\lambda_i \Delta t}  \int_{t}^{t+ \Delta t} e^{-\lambda_i (t -t')} n(t') dt'
+C(0) e^{-\lambda_i t} e^{-\lambda_i \Delta t} 
\end{displaymath}
\begin{equation}
= C(t) e^{-\lambda_i \Delta t} + \delta C_i(t,\Delta t) \, ,
\end{equation}
wobei
\begin{equation}
\delta C_i (t , \Delta t) =
\frac{b_i}{\Lambda^*} e^{-\lambda_i \Delta t} \int_{t}^{t+\Delta t} e^{-\lambda_i (t-t')} n(t') dt'
\end{equation}
durch die Annahme approximiert werden soll, dass die Neutronenzahl (gemessen) zur Zeit
$t$ und $t+\Delta t$ linear interpoliert werden kann durch
\begin{equation}
n(t') = n(t) + \frac{t'-t}{\Delta t} (n(t + \Delta t) - n(t)) \, , \quad t' \in [t, t+ \Delta t ] \, .
\end{equation}
Damit ergibt sich
\begin{equation}
\delta C_i (t , \Delta t) =
\frac{b_i}{\Lambda^*} e^{-\lambda_i \Delta t} \int_{t}^{t+\Delta t} e^{-\lambda_i (t-t')} \biggl[
n(t) +  \frac{t'-t}{\Delta t} (n(t + \Delta t) - n(t)) \biggr] dt' \, .  \label{int}
\end{equation}
Diskretisiert man nun die Zeitvariable gem\"ass
$t^k = k \Delta t$, $k \in \mathds{N}_0$ and setzt
$n^k=n(t^k)$, $C_i^k=C_i( t^k)$, so kann Gleichung (\ref{int}) in der Form
($\tilde{t} = t'-t$)
\begin{displaymath}
\delta C_i (t^{k-1} , \Delta t) =
\frac{b_i}{\Lambda^*} e^{-\lambda_i \Delta t} \int_{0}^{\Delta t} e^{\lambda_i \tilde{t}} \biggl[
n^{k-1} +  \frac{\tilde{t}}{\Delta t} (n^k - n^{k-1}) \biggr] d\tilde{t} 
\end{displaymath}
\begin{displaymath}
= \frac{b_i}{\Lambda^* \lambda_i} \Biggl \{ \Biggl( n^{k-1} - \frac{n^k - n^{k-1}}{\lambda_i \Delta t} \Biggl)
\bigl( 1 - e^{-\lambda_i \Delta t} \bigr) + n^k - n^{k-1} \Biggr\} 
\end{displaymath}
\begin{equation}
= \frac{b_i}{\Lambda^* \lambda_i^2 \Delta t}  \bigl( (n^k - n^{k-1})
(e^{-\lambda_i \Delta t} -1) +  (n^k - n^{k-1} e^{-\lambda_i \Delta t}) \lambda_i \Delta t  \bigr)
\end{equation}
geschrieben werden, 
wobei die Integrationsformeln
\begin{equation}
\int \tilde{t} e^{\lambda \tilde{t}} d\tilde{t} = \frac{1}{\lambda^2} ( \lambda \tilde{t} -1)
e^{\lambda \tilde{t}}  + konst.  \, ,
\quad \int_0^{\Delta t} \tilde{t} e^{\lambda \tilde{t}} d \tilde{t}  =
\frac{1}{\lambda^2} (\lambda \Delta t e^{\lambda \Delta t} - e^{\lambda \Delta t} + 1)
\end{equation}
und
\begin{equation}
\int_0^{\Delta t}  e^{\lambda \tilde{t}} d\tilde{t} = \frac{1}{\lambda} (e^{\lambda \Delta t} -1 )
\end{equation}
verwendet wurden.
Damit l\"asst sich die Rekursionsgleichung zur Berechnung des Anteils verz\"ogerter Neutronen angeben
\begin{equation}
C_i^k = C_i^{k-1} e^{-\lambda_i \Delta t} +
 \frac{b_i}{\Lambda^* \lambda_i} \Biggl \{ \Biggl( n^{k-1} - \frac{n^k - n^{k-1}}{\lambda_i \Delta t} \Biggl)
\bigl( 1 - e^{-\lambda_i \Delta t} \bigr) + n^k - n^{k-1} \Biggr\} \, ,
\end{equation}
und die Reaktivit\"at eines Reaktors zum aktuellen Zeitpunkt $t^k$ ergibt sich aus zeitlich
regelm\"assig erfolgten Neutronenflussmessungen $\Phi^k \sim n^k$, $k=0,1,\ldots,k$, zu
\begin{equation}
\rho_*^k = 1 + \frac{\Lambda^*}{n^k} \Biggl\{ \frac{n^k - n^{k-1}}{\Delta t}
- \sum_{i=1}^6 \lambda_i C_i^k (t) \Biggr\} 
\end{equation}
oder
\begin{equation}
\rho_*^k = 1 + \frac{\Lambda^*}{\Phi^k} \Biggl\{ \frac{\Phi^k - \Phi^{k-1}}{\Delta t}
- \sum_{i=1}^6 \lambda_i C_i^k (t) \Biggr\} \, ,
\end{equation}
wenn die $C_i^k$ ebenfalls aus den $\Phi^k$ anstelle der $n^k$ berechnet wurden.

\subsection*{Die Sechsfaktorformel}
Die Sechsfaktorformel geht in vereinfachter Form auch als Neutronen-Strahlungsverluste (Le\-cka\-ge) ver\-nach\-l\"as\-sigen\-de
Vierfaktorformel auf Eugene Wigner \cite{Wigner}
und Enrico Fermi \cite{Fermi} zur\"uck und veranschaulicht die Bedeutung des Multiplikationsfaktors $k$
aus Gleichung (\ref{multifac}). Die Formel beschreibt die Anzahl der {\emph{thermischen} Neutronen, die durch ein im Brennstoff
eines Reaktors absorbiertes Neutron erzeugt werden und wieder den Weg zur\"uck in den Brennstoff finden,
wo sie schliesslich erneut absorbiert werden. Die Formel lautet
\begin{equation}
k= \eta \cdot \epsilon \cdot P_s \cdot p \cdot P_{th} \cdot f \, .
\end{equation}
Absorbiert also der Brennstoff eines Reaktors $n$ thermische Neutronen, so generieren diese letzlich $k \cdot n$
direkte thermische Nachfahren, welche wiederum vom Brennstoff absorbiert werden und nicht anderswo verlustig gehen.
Dabei beschreibt $\eta$ (thermal fission factor) die sogenannte \emph{Neutronenergiebigkeit} des Brennstoffs. Nicht jedes
thermische Neutron, welches im
Brennstoff eines Reaktors absorbiert wird, l\"ost aber eine Spaltung aus. \"Uber ein F\"unftel der thermischen Neutronen
wird in angereichertem Uran (3\% $^{235}$U, 97\% $^{238}$U) eingefangen, ohne eine Spaltung auszul\"osen.
Dabei entstehen die Isotope $^{236}$U und $^{239}$U, wobei das $^{239}$U zuerst in $^{239}$Np und schliesslich zu $^{239}$Pu
zerf\"allt. Somit setzt ein im Brennstoff eines Druck- oder Siedewasserreaktors
absorbiertes Neutron im Mittel etwas weniger als zwei neue thermische Neutronen frei, obwohl bei einer Kernspaltung etwa
$\nu \simeq 2.5$ Neutronen entstehen. Es ist also zwischen den Begriffen Kern{\emph{brenn}stoff und \emph{Spalt}stoff zu unterscheiden.
Zum Brennstoff z\"ahlt man typischerweise alle Uran- und Plutoniumisotope, welche in einem Brennelement enthalten sind,
w\"ahrend die relevanten Spaltstoffe durch $^{235}$U, $^{239}$U und $^{241}$Pu repr\"asentiert werden (nat\"urlich k\"onnen
in dieser Betrachtung auch etwas unkonventionellere Isotope wie thermisch spaltbares $^{233}$U oder Thoriumisotope etc. mit einbezogen werden).
Es ist zu erw\"ahnen, dass auch ein erfolgreiches Neutron, welches eine Kernspaltung ausl\"ost, als absorbiert gilt: es wird zuerst
vom getroffenen Kern einverleibt, bevor dieser nach einer kurzen angeregten Phase als Compoundkern zerplatzt und $\nu$ Neutronen
freisetzt.\\

\noindent Die bei einer Kernspaltung entstehenden hochenergetischen Neutronen k\"onnen allerdings mit geringer
Wahr\-schein\-lich\-keit auch $^{238}$U-Kerne spalten; dadurch erh\"oht sich die Anzahl der bereits vorhandenen
Neutronen um einen \emph{Schnell\-spaltfaktor} $\epsilon$ (fast fission factor). Zugleich entkommen einige der schnellen Neutronen
mit der \emph{schnellen Ent\-komm\-wahr\-schein\-lich\-keit} dem Reaktor und sind f\"ur die Kettenreaktion f\"ur immer
verloren. Der \emph{schnelle Verbleibfaktor} $P_{s}$ (fast non-leakage probability) tr\"agt diesem Umstand Rechnung.
\\

\noindent Beim anschliessenden Moderationsprozess verlieren die Neutronen durch
St\"osse (beispielsweise an Moderator\-kernen wie $^{12}$C in Graphit, $^2$H in schwerem Wasser oder $^1$H
in leichtem Wasser) an kinetischer Energie.
Im Energiebereich von 7eV...200eV
laufen sie dabei Gefahr, durch sogenannte Resonanz\-absorption durch $^{238}$U-Kerne eingefangen zu
werden. Wiederum entsteht bei diesem Prozess $^{239}$U, welches sich in Plutonium umwandeln wird.
Es kommt nur ein gewisser Anteil der Neutronen im thermischen Bereich an, dieser ist gegeben
durch die \emph{Resonanz\-ent\-komm\-wahr\-scheinlich\-keit} $p$ (resonance escape probability).
In einem Siedewasserreaktor der Gigawatt-Klasse liegt die \"Uber\-le\-bens\-wahr\-schein\-lich\-keit eines Neutrons
beim Abbremsvorgang beispielsweise bei ca. $p \simeq 0.8=80\%$.
Allerdings h\"angt $p$ in erster Linie vom Wassergehalt im Reaktorkern ab; dieser kann wie bereits erw\"ahnt durch Variation des
Kernmassenstroms, also der durch den Reaktorkern pro Zeiteinheit gef\"orderten Wassermenge, kontrolliert
werden, was wiederum eine unmittelbare Steuerung der Kettenreaktion erm\"oglicht.
Die Moderationsf\"ahigkeit eines thermischen Reaktors ist definiert durch das Produkt $W_{th}= \epsilon \cdot P_{s} \cdot p \simeq p$.\\

\noindent Die thermalisierten Neutronen diffundieren schliesslich durch den Reaktorkern. W\"ahrend der Abbremsvorgang
in einem Leichtwasserreaktor normalerweise innerhalb von einigen $10^{-6}$s abgeschlossen ist, lebt ein thermisches
Neutron anschliessend mit etwa $10^{-4}$s noch vergleichsweise lange. Der \emph{thermische
Verbleibfaktor} $P_{th}$ (thermal non-leakage probability) beschreibt den Anteil der Neutronen, die w\"ahrend dieser Phase
nicht aus dem Reaktorkern hinaus\-diffun\-dieren.
Die \emph{thermische Nutzung} $f$ (thermal utilization factor) schliesslich beschreibt den Anteil der thermischen Neutronen, welche
wieder im Brennstoff und nicht im restlichen Material der Reaktorkerns absorbiert werden. Mit einem f\"ur Leichtwasserreaktoren
typischen Wert von $f \simeq 75 \%$ werden also rund $25\%$ der thermischen Neutronen durch Materialien im Reaktorkern
absorbiert, welche nicht zum Brennstoff gez\"ahlt werden.
Viele der thermisch absorbierten Neutronen werden durch die Protonen im Wasser absorbiert, welche sich mit den Neutronen zu Deuteronen verbinden.
Eine \"ahnlich wichtige Rolle spielt in Druckwasserreaktoren die Absorption durch im K\"uhlwasser in Form von Bors\"aure (B(OH)$_3$)
gel\"ostem Bor ($^{10}$B).
Schliesslich bilden sich im Reaktorbetrieb Spaltprodukte, welche die thermische Nutzung massgeblich beeinflussen.
Die wichtigste und unter Umst\"anden zeitlich stark variierende Rolle spielt dabei das instabile Isotop $^{135}$Xe,
von geringerer Bedeutung ist $^{149}$Sm.  \\

\noindent Realistische Werte f\"ur einen quasi kritischen Leistungsreaktor bei konstanter
Nominalleistung w\"aren in obiger Reihenfolge
\begin{equation}
k = 1 = 1.69 \cdot 1.04 \cdot 0.96 \cdot 0.787 \cdot 0.996 \cdot 0.756 \, .
\end{equation}

\noindent  Es vergr\"ossert sich also die thermische Neutronenzahl in einem \"uberkritischen Reaktor
beim Durchlaufen eines oben beschriebenen Neutronenzyklus oder
\emph{"Fermizyklus"} um den sogenannten Multiplikationsfaktor $k$, den wir nun mit $k$ aus
Gleichung (\ref{multifac}) identifizieren wollen. Man mag sich n\"amlich fragen, wie lange sich ein Neutron durchschnittlich im
Neutronenzyklus aufh\"alt. Im Rahmen der Eingruppengleichung betrachten
wir hierzu einen schwach \"uberkritischen Reaktor der Reaktivit\"at $\rho \ll \beta$, in welchem
die Neutronenzahl mit stabiler Reaktorperiode zeitlich exponentiell w\"achst.
Dann gilt in guter N\"aherung f\"ur die reziproke Reaktor\-periode gem\"ass Gleichung (\ref{naefo})
\begin{equation}
\omega \simeq \frac{\lambda \rho}{\beta - \rho} \simeq \frac{\lambda \rho}{\beta} \, .
\end{equation}
Innerhalb der Zeit $t_F$ soll sich also die Neutronenzahl um den Faktor $k$ erh\"ohen.
Mit einer stabilen Reaktor\-periode gilt also wegen $e^{x} \simeq 1+x$ f\"ur $|x| \ll 1$
\begin{equation}
k=e^{\omega \cdot t_F} \simeq 1+ \omega \cdot t_F = \frac{1}{1-\rho} \simeq 1+\rho \, ,
\end{equation}
also ist
\begin{equation}
t_F \simeq \frac{\rho}{\omega} \simeq \frac{\beta}{\lambda} = \beta \cdot l \equiv L \, . \label{fermitime}
\end{equation}
In Gleichung (\ref{fermitime}) taucht die mittlere Lebensdauer $l \simeq 13$s der Vorl\"aufer auf. Diese dominiert
die mittlere Dauer, die ein Fermizyklus in Anspruch nimmt. Zwar durchl\"auft die Mehrzahl der
Neutronen den Fermizyklus innerhalb einer kurzen Zeit der Gr\"ossenordnung der Neutronenlebensdauer oder
der Generationszeit $\Lambda \simeq \tau$.
Der geringe Anteil von weniger als einem Prozent der Neutronen, welche in Vorl\"aufern gespeichert
den Fermizyklus verz\"ogert, lebt aber ungleich l\"anger und bestimmt im verz\"ogert \"uberkritischen
Reaktor die mittlere Verweilzeit der Neutronen im Neutronenzyklus, bei der sich die Neutronenzahl um den
Multiplikationsfaktor $k=\tau/\Lambda$ erh\"oht.
\\

\noindent Man nennt die durch die verz\"ogerten Neutronen dominierte mittlere Generationszeit $\bar{\Lambda}$ der Neutronen im
leicht \"uberkritischen Reaktor $L=(1-\beta) \Lambda + \beta \cdot (l+\Lambda) \simeq \beta \cdot l = \bar{\Lambda} = 0.083$s
(f\"ur $^{235}$U) auch verz\"ogerte Lebensdauer. Betrachten wir einen Reaktor mit Multiplikationsfaktor $k=1.0005$ beziehungsweise
$\rho \simeq 50$pcm, so verdoppelt sich in einem solchen Reaktor die Neutronenzahl nach 1387 mittleren
Neutronenzyklen, da gilt $1.0005^{1387} = 2$. Die Verdoppelungszeit betr\"agt also $\rm 1387 \cdot 0.083s =
115s$, in recht passabler \"Ubereinstimmung mit dem Wert $t_2 = 90$s aus der Inhourkurve (Figur 2).
In einem mit reinem $^{239}$Pu betriebenen Reaktor w\"are $L_{Pu-239}= 0.03\rm s$ deutlich geringer.
\\

\noindent Bei geringen Reaktivit\"aten w\"achst die Neutronenzahl in einem Reaktor also gem\"ass der N\"aherung
\begin{equation}
 n(t) \simeq n_0 \cdot \exp \biggl( \frac{\rho}{L} \cdot t \biggr) \simeq
n_0 \cdot \exp \biggl( \frac{k-1}{L} \cdot t \biggr)  \label{compact_formula}
\end{equation}
mit einer geeignet gew\"ahlten Startzahl $n_0$ zum Zeitpunkt $t=0$, nachdem sich der Reaktor auf eine stabile Reaktorperiode
eingeschwungen hat. Gleichung (\ref{compact_formula}) kann auch folgendermassen motiviert werden:
Im \"uberkritischen, aber sicheren prompt unterkritischen Bereich hat ein einzelner Neutronenzyklus unter Be\-r\"uck\-sichtigung
der verz\"ogerten Neutronen die Dauer $\beta \cdot l$. Nach einer Zeit $t$ betr\"agt demnach die Neutronenzahl
\begin{equation}
n(t)=n_0 \cdot k^{t/L} = n_0 \cdot \exp \biggl( \frac{t}{L} \ln(k) \biggr) \simeq n_0 \cdot \exp \biggl( \frac{t}{L}
\ln(1 + \rho) \biggr) \simeq n_0 \cdot \exp \biggl( \frac{\rho}{L} \cdot t \biggl) = n_0 \cdot e^{\frac{\lambda \rho}{\beta}\cdot t} \, .
\end{equation}

\subsection*{Diffusion}
Die Neutronenflussdichte $\Phi_{th}$ in einem Reaktorkern ist eine r\"aumlich variierende Gr\"osse.
Typische Werte der Neutronenflussdichten in Leistungreaktoren bewegen sich im Bereich von einigen
$10^{13} \rm cm^{-2} s^{-1}$.
R\"aumliche Gradienten der Flussdichte $\vec{\nabla} \Phi_{th}$ f\"uhren zum lokalen Transport thermischer Neutronen,
welcher durch die Neutronenstromdichte $\vec{j}_{th}(\vec{r})$ im Rahmen der Diffusionsgleichung
($\bar{v}$ sei konstant)
\begin{equation}
\vec{j}_{th}= -D \cdot \vec{\nabla} \Phi_{th}=-D \cdot grad \, \Phi_{th} = -D \cdot \bar{v} \cdot grad \,
n_{th}^{_\Box} \label{diffu1}
\end{equation}
beschrieben werden kann. $D$ spielt hierbei die Rolle einer Diffusionskonstanten mit der Dimension
einer L\"ange. Der Neutronendiffusionsstrom trachtet nach einem Dichteausgleich durch den Transport
von Neutronen aus Regionen h\"oherer Dichte in Regionen niedrigerer Dichte.
Die lokalen Neutronendichteschwankungen durch reine Transportph\"anomene (d.h. bei unrealistischer
Vernachl\"assigung der Neutronenproduktion durch fortw\"ahrende Kernspaltungen)
k\"onnen
mit Hilfe der Kontinuit\"atsgleichung
\begin{equation}
\frac{\partial n_{th}^{_\Box}}{\partial t}= -\vec{\nabla} \cdot \vec{j}_{th} =- div \, \vec{j}_{th} \label{conti}
\end{equation}
beschrieben werden. Aus der Kombination von Gleichung (\ref{diffu1}) und (\ref{conti}) ergibt sich
die Diffusionsgleichung (\emph{heat equation})
\begin{equation}
\frac{\partial n_{th}^{_\Box}}{\partial t} = D \cdot \Delta \Phi_{th}
\end{equation}
oder
\begin{equation}
\frac{\partial n_{th}^{_\Box}}{\partial t} = \tilde{D} \cdot \Delta n_{th}^{_\Box} \, ,  \quad \, 
\tilde{D}=D \cdot \bar{v} \, .
\end{equation}
Die \emph{Green}sche Funktion (oder der \emph{Kern}, die Fundamentall\"osung) dieser Gleichung im homogenen
Raum mit konstanter Diffusionskonstante lautet
\begin{equation}
H(\vec{r}) )= \frac{1}{(4 \pi \tilde{D} t)^{3/2}}  e^{-\frac{\vec{r}^2}{4 \tilde{D} t}} \quad . \label{kernel}
\end{equation}
Sie beschreibt das diffusive Zerfliessen einer zur Zeit $t=0$ punktf\"ormigen St\"orung der
Neutronendichte. Die Zeitskala, mit der die Gausssche Glockenfunktion in Gleichung (\ref{kernel}) 
zerfliesst, ist gegeben durch
$t_D \simeq \frac{R^2}{4 D}$, wenn $R$ eine charakteristische Ausdehnung der betrachteten Konfiguration
beschreibt. F\"ur $D \simeq 1$cm, $R \simeq 4$m und $\bar{v} = 3000$m/s ergibt sich $t_D \simeq
0.1$s - damit gleichen sich Schwankungen der Neutronendichte in einem prompt unterkritischen Reaktor
relativ schnell gegen\"uber der stabilen Reaktorperiode aus, doch im prompt (\"uber)kritischen Reaktor
hat das Punktmodell seine G\"ultigkeitsgrenzen (auch aus anderen Gr\"unden) erreicht.

%------------------------------------------------------------------------------------------------------------------------------------

\newpage

\subsection*{Anhang}

\subsection*{Daten der sechs Vorl\"aufergruppen von $^{235}$U}
Die Daten f\"ur $^{235}$U stammen aus einer Arbeit von Keepin, Wimett und Zeigler \cite{Keepin}.
\begin{table}[h]
\begin{center}
\begin{tabular}{|c|c|c|c|c|c|}
\hline
& & & & & \\
Gruppe & Vorl\"aufernuklide  &  Relativanteil  & Zerfallskonstante $\lambda_i$ &
Lebensdauer $l_i$ & Halbwertszeit $t_{1/2,i}$ \\
  &   &  $\beta_i/\beta$  & [s$^{-1}$] & [s]  &  [s]\\
& & & & & \\
\hline
\hline
& & & & & \\
1    &     $^{93}$Br, ...    &     0.042   & 3.0137   &    0.332    & 0.230  \\

%\hline
& & & & & \\
2    &     $^{140}$I, $^{145}$Cs...    &     0.116   & 1.1360   &    0.880    & 0.610  \\

%\hline
& & & & & \\
3    &     $^{144}$Cs, $^{139}$I, $^{90}$Br, ...    &     0.396   & 0.3014   &    3.318    & 2.300  \\

%\hline
& & & & & \\
4    &     $^{138}$I, $^{89}$Br, ...    &     0.195   & 0.1114   &    8.974    & 6.222  \\

%\hline
& & & & & \\
5    &     $^{137}$I, $^{88}$Br, ...    &     0.219   & 0.03051   &    32.778    & 22.72  \\

%\hline
& & & & & \\
6    &     $^{87}$Br    &     0.032   & 0.01244   &    80.387    &  55.72  \\
& & & & & \\
\hline
\end{tabular}
\end{center}
\end{table}

\vskip -0.4cm
\noindent Zur Illustration:
Es ist $t_{1/2}^{^{88}{\rm{Br}}}=16.3$s, $t_{1/2}^{^{137}}{\rm{I}}=24.2$s. Es gilt
$\sum \limits_{i=1}^{6} \beta_i = \beta \simeq 0.0064$,
$l=\sum \limits_{i=1}^{6} \beta_i l_i \simeq 12.93$s.\\

\noindent Mittlere Neutronenenergien $\bar{E}_i$ der $i$-ten Vorl\"aufergruppe sind $\bar{E}_2=0.42$MeV, $\bar{E}_3=0.62$MeV, 
$\bar{E}_4=0.43$MeV, $\bar{E}_5=0.56$MeV und $\bar{E}_6=0.25$MeV.\\

\noindent Die Anteile verz\"ogerter Neutronen f\"ur weitere Spaltkerne sind gegeben durch:\\
$\beta_{U-238} = 1.57\%$,
$\beta_{Th-232} = 2.2\%$, $\beta_{U-233} = 0.27\%$, $\beta_{Pu-239} = 0.21\%$, $\beta_{Pu-241}=0.54\%$,
$\beta_{Am-241}=0.13\%$, $\beta_{Am-241}=0.24\%$.

\subsection*{Das Spektrum der prompten Spaltneutronen}
Nach Watt \cite{Watt} l\"asst sich das kinetische Energiespektrum der bei der Kernspaltung prompt
freigesetzten Neutronen durch den Ausdruck
\begin{equation}
\chi(E)=0.484 \sinh \sqrt{2 E} e^{-E} \label{Watt}
\end{equation}
parametrisieren. Dabei wird die Neutronenenergie $E$ in MeV gemessen.
In ein infinitesimales Energieintervall werden dann je Spaltung $\bar{\nu}(E_f) \cdot \chi(E) dE$ Neutronen
erzeugt, wobei die Neutronenausbeute $\bar{\nu}$ selbst von der Energie der Neutronen abh\"angt,
welche die Spaltung ausl\"osen. Neutronen, die mit hoher Energie eine Spaltung ausl\"osen, setzen auch
eine gr\"ossere Anzahl Spaltneutronen frei. Dieser Zusammenhang kann durch eine lineare Beziehung
\begin{equation}
\bar{\nu}(E_f) = \nu_0 + a \cdot E_f
\end{equation}
dargestellt werden und wurde experimentell im Energiebereich bis 15MeV untersucht (\cite{Keepbook},
siehe nachfolgende Tabelle).\\

\noindent In guter N\"aherung l\"asst sich das thermische Spaltneutronenspektrum auch durch eine
Maxwell-Boltzmann-Verteilung darstellen:
\begin{equation}
\chi^{MB}(E)=\frac{2}{\pi^{1/2} T^{3/2}} E^{1/2} e^{-E/T} \, , \label{Maxw}
\end{equation}
mit $T=1.29$MeV f\"ur $^{235}$U. Eine approximative Maxwellverteilung der emittierten Neutronen
l\"asst die Vorstellung eines Nukleonengases in einem hochangeregten Kern zu, der  bei einer Temperatur
von $1.5 \cdot 10^{15}$K Neutronen abdampft.
Die Verteilung $\chi^{MB}(E)$ in Gleichung (\ref{Maxw}) ist gem\"ass
\begin{equation}
\int \limits_{0}^{\infty} \chi^{MB}(E) dE =1
\end{equation}
normiert.

\begin{table}[h]
\begin{center}
\begin{tabular}{|c|c|c|c|}
\hline
& & &  \\
Kernbrennstoff & $\nu_0$  &  a  & Energiebereich  \\
  &   & [MeV$^{-1}$]  & [MeV]\\
& & &  \\
\hline
\hline
& & &  \\
$^{232}$Th    &      1.87  &     0.164   & 0$<$E$<$15    \\

\hline
& & &  \\
$^{233}$U    &      2.48  &     0.075   & 0$<$E$<$1    \\

%\hline
& & &  \\
   &      2.41  &     0.136   & 1$<$E$<$15    \\

\hline
& & &  \\
$^{235}$U    &      2.43  &     0.065   & 0$<$E$<$1    \\

%\hline
& & &  \\
   &      2.35  &     0.150   & 1$<$E$<$15    \\

\hline
& & &  \\
$^{238}$U    &      2.30  &     0.160   & 0$<$E$<$15    \\

\hline
& & &  \\
$^{239}$Pu    &      2.87  &     0.138   & 0$<$E$<$1    \\

%\hline
& & &  \\
   &      2.91  &     0.133   & 1$<$E$<$15    \\

& & &  \\
\hline
\end{tabular}
\end{center}
\end{table}

\noindent Der Parameter $T=1.29$MeV ist mit der mittleren Energie der Spaltneutronen \"uber die Beziehung
\begin{equation}
\bar{E}=\int \limits_{0}^{\infty} E \chi^{MB}(E) dE = \frac{3}{2} T
\end{equation}
verkn\"upft.
Die mittleren Neutronenenergien sind f\"ur die relevanten Spaltstoffe im Wesentlichen gleich: F\"ur obiges $T$
ist $\bar{E}_{U-235}=1.935 \pm 0.05$MeV, weiter ist $\bar{E}_{U-233}=1.96 \pm 0.05$MeV und
$\bar{E}_{Pu-239}=2.00 \pm 0.05$MeV \cite{Keepbook}.

\begin{center}
\includegraphics[width=8.0cm, angle=270]{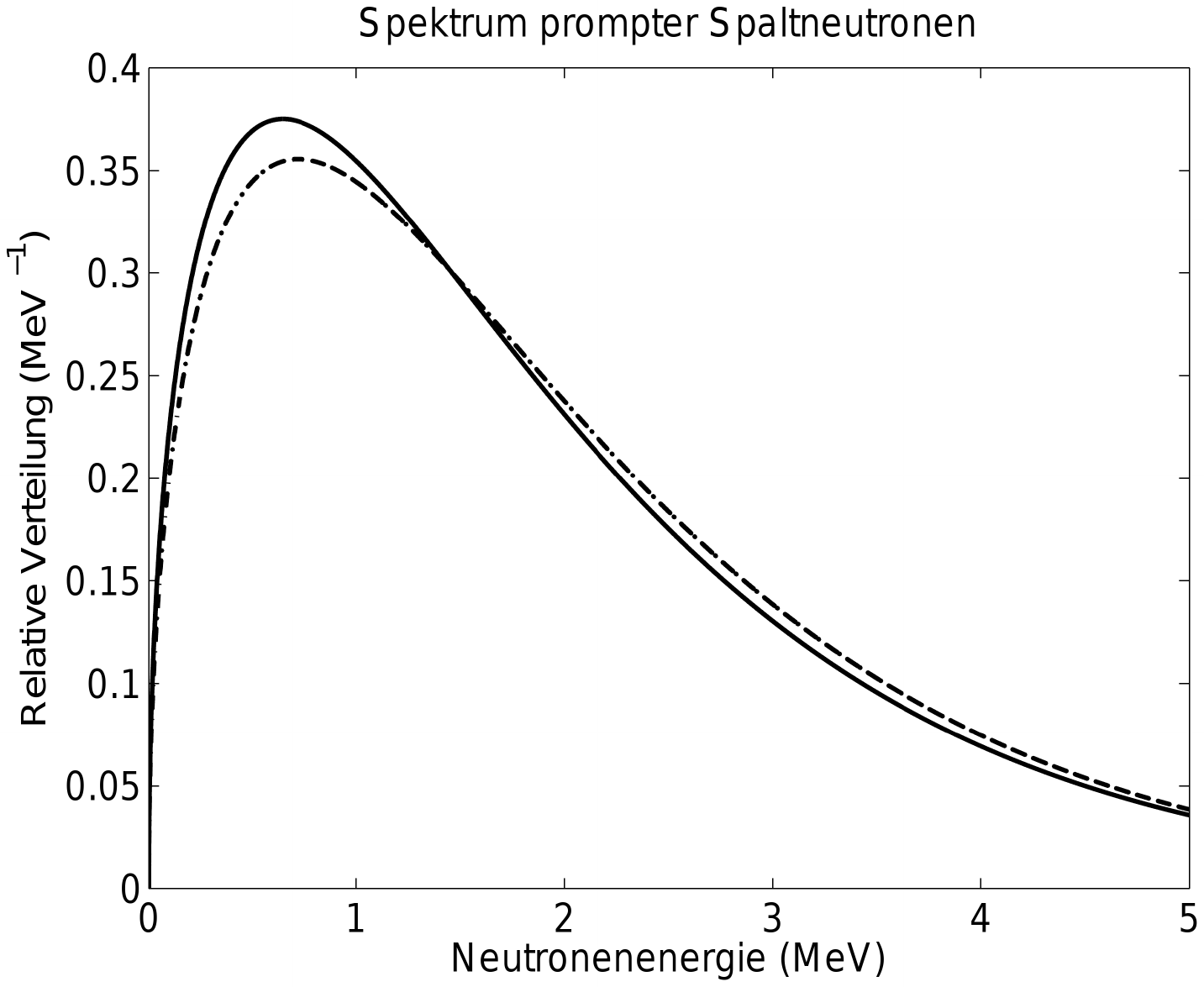}\
\end{center}
Figur 4: Spektrum der Spaltneutronen nach Gleichung (\ref{Watt}) (ausgezogene Kurve)
und Gleichung (\ref{Maxw}) (punktgetrichene Kurve).

%\newpage

\subsection*{L\"osung der Inhour-Gleichung}
Die reaktorkinetischen Grundgleichungen mit $m$ Vorl\"aufergruppen lauten
\begin{equation}
\frac{dn}{dt}  =  \frac{\rho-\beta}{\Lambda} n + \lambda_1 C_1 + \ldots +\lambda_m C_m \, , \label{neutron}
\end{equation}
\begin{equation}
\frac{dC_i}{dt} =   \frac{\beta_i}{\Lambda} n - \lambda_i  C_i \, , \quad  i=1\ldots m \, . \label{precursors}
\end{equation}
Die Summe der Gleichungen (\ref{neutron}) und (\ref{precursors}) f\"uhrt auf
\begin{equation}
\dot{n} + \dot{C}_1 + \ldots + \dot{C}_m = \frac{\rho}{\Lambda} n. \label{summa}
\end{equation}
Mit dem Exponential-Ansatz 
\begin{equation}
n(t) = n_0 \cdot e^{\omega t} \, ,
\end{equation}
\begin{equation}
C_i(t)= C_{i,o} \cdot e^{\omega t}
\end{equation}
f\"uhrt Gleichung (\ref{precursors}) auf
\begin{equation}
\omega C_{i,o}= \frac{\beta_i}{\Lambda} n_0 - \lambda_i C_{i,0} \quad \mbox{oder} \quad C_{i,0}=\frac{\beta_i}
{\Lambda ( \omega+\lambda_i)} n_0 \, .
\end{equation}
Dieses Zwischenresultat ergibt eingesetzt in Gleichung (\ref{summa}) die ber\"uhmte Inhour-Gleichung
\begin{equation}
\omega + \sum \limits_{i=1}^{m} \frac{\omega \beta_i}{\Lambda (\omega+\lambda_i)} = \frac{\rho}{\Lambda}
\end{equation}
oder
\begin{equation}
\omega \Biggl( \Lambda +  \sum \limits_{i=1}^{m} \frac{\beta_i}{\omega+\lambda_i} \Biggr) = \rho \, . \label{inho}
\end{equation}
Diese Gleichung ist \"aquivalent zu einer polynomialen Gleichung m+1-ten Grades. Durch Multiplikation der
Inhour-Gleichung mit $(\omega+\lambda_1) \cdot \ldots \cdot (\omega+\lambda_m)$ ergibt sich n\"amlich
\begin{equation}
\omega \Lambda (\omega+\lambda_1) \cdot \ldots \cdot (\omega+\lambda_m)+\omega
\sum \limits_{i=1}^{m} \beta_i (\omega+\lambda_1) \cdot \ldots \cdot \xcancel{(\omega+\lambda_i)} \cdot \ldots
 \cdot (\omega+\lambda_m)= \rho (\omega+\lambda_1) \cdot \ldots \cdot (\omega+\lambda_m) \, ,
\end{equation}
was kompakter in der Form
%\begin{equation}
%\omega \Lambda \prod_{i=1}^{m} (\omega+\lambda_i)+ \omega \sum_{i=1}^{m} \beta_i \prod_{\substack{j=1\\  j \neq %i}}^{m}
%(\omega + \lambda_j) = \rho \prod_{i=1}^{m} (\omega+\lambda_i)
%\end{equation}
\begin{equation}
(\omega \Lambda - \rho) \prod_{i=1}^{m} (\omega+\lambda_i)+ \omega \sum_{i=1}^{m} \beta_i \prod_{\substack{j=1\\  j \neq i}}^{m}
(\omega + \lambda_j) = 0
\end{equation}
geschrieben werden kann.
Es existieren also $m+1$ reziproke (oder inverse) Reaktorperioden $\omega_1 \geq \ldots \geq \omega_{m+1}$, welche die obige Gleichung
l\"osen, wobei wir in der Praxis vom nicht-entarteten Fall ausgehen k\"onnen und annehmen,
dass die reziproken Reaktorperioden
paarweise verschieden sind ($\omega_i \neq \omega_j$ f\"ur $i \neq j$).
Trivialerweise gilt $\lambda_i \neq \lambda_j$ f\"ur $i \neq j$, da man verschiedene Vorl\"aufer mit identischer mittlerer
Lebensdauer in einer gemeinsamen Vorl\"aufergruppe zusammenfassen w\"urde.\\
 
\noindent Es ist nun leicht zu zeigen, dass f\"ur positive Reaktivit\"at $\rho>0$ genau eine positive reziproke stabile
Reaktorperiode $\omega_1$ existiert, welche die Bedingung (\ref{inho}) 
\begin{equation}
\omega_1 =  \frac{\rho}{ \Lambda +  \sum \limits_{i=1}^{m} \frac{\beta_i}{\omega_1+\lambda_i}}
\end{equation}
erf\"ullt. Betrachten wir n\"amlich im betrachteten Falle
die f\"ur $\omega>0$ offensichtlich streng monoton wachsende, konkave und beschr\"ankte Funktion
\begin{equation}
f(\omega)=\frac{\rho}{ \Lambda +  \sum \limits_{i=1}^{m} \frac{\beta_i}{\omega+\lambda_i}} \, , \label{funktion}
\end{equation}
so gilt offensichtlich
\begin{equation}
0<f(0)=\frac{\rho}{\Lambda + \sum \limits_{0}^{m} \frac{\beta_i}{\lambda_i}} = \frac{\rho}{\Lambda+\beta l} <
\lim_{\omega \rightarrow \infty} f(\omega) = \frac{\rho}{\Lambda} < \infty \quad .
\end{equation}
Wie in Figur 5 dargestellt liegen f\"ur $\omega \geq 0$ die Funktionswerte von $f(\omega)$ im Intervall
$[ \frac{\rho}{\Lambda+\beta l},  \frac{\rho}{\Lambda}]$ ($l$ ist die mittlere Lebensdauer der Vorl\"aufer),
sodass aus den Eigenschaften des Funktionsgraphen von $f$ zwangsl\"aufig
die Existenz einer eindeutigen positiven Reaktorperiode $1/\omega_1$ mit $\omega_1= f(\omega_1)$ folgt.
\begin{center}
\includegraphics[width=10.8cm,angle=-90]{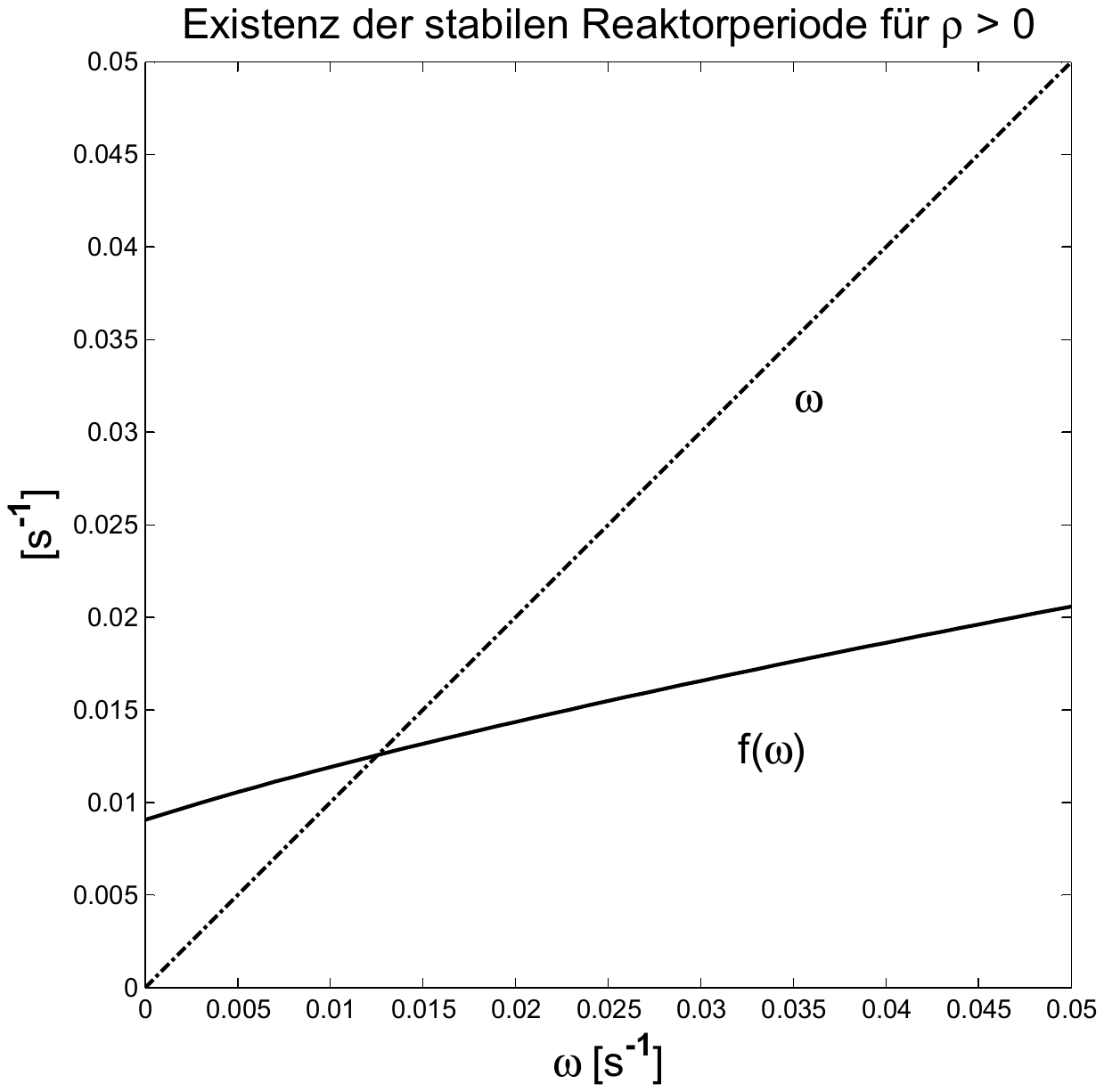}\\
\vspace{0.3cm} Figur 5: Verlauf der gem\"ass Gleichung (\ref{funktion}) definierten Funktion $f(\omega$)
f\"ur einen Beispielwert $\rho = 0.00075 = 75$pcm.
\end{center}

\noindent Aus der obigen Berechnung folgt eine einfache iterative Berechnungsmethode f\"ur die stabile Reaktorperiode.
Als vern\"unftigen Startwert w\"ahle man $\omega_1^1=\frac{\rho}{\beta l}$.
Dann berechne man Folgewerte $\omega_1^2,$ $\omega_1^3$, $\ldots$ gem\"ass der Vorschrift

\begin{equation}
\omega_1^{k+1}=\frac{\rho}{ \Lambda +  \sum \limits_{i=1}^{m} \frac{\beta_i}{\omega_1^k+\lambda_i}}
\, . \label{funktion}
\end{equation}
Die so erhaltene Zahlenfolge $\{\omega_1^k\}_{k \in \mathds{N}}$ konvergiert rasch gegen $\omega_1$:
\begin{equation}
\lim_{k \rightarrow \infty} \omega_1^k = \omega_1.
\end{equation}

\begin{center}
\includegraphics[width=14.8cm]{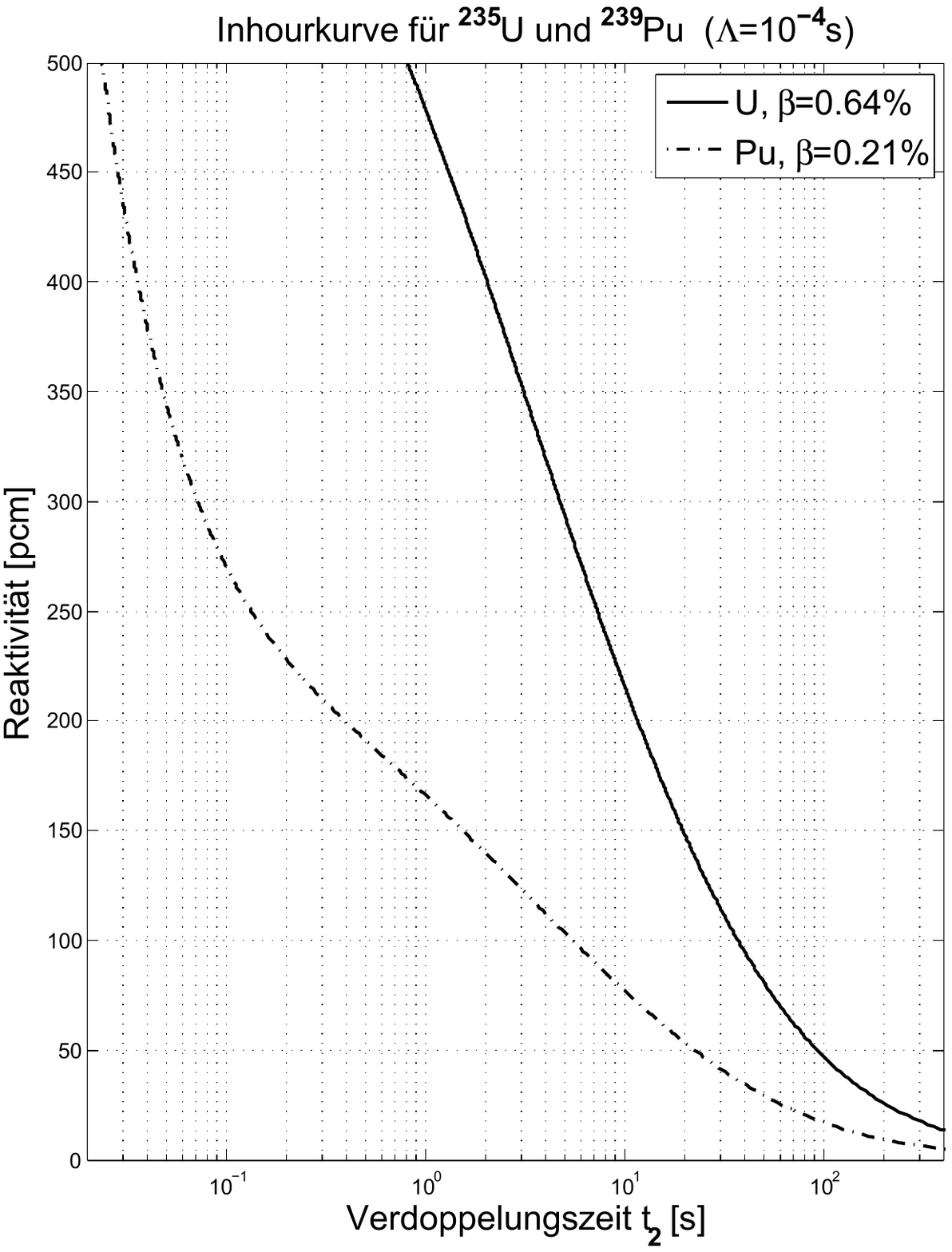}\\
\vspace{-0.0cm} Figur 6: Verdoppelungszeit versus Reaktivit\"at in der Sechsgruppenn\"aherung f\"ur
$^{235}$U mit $\beta_{U-235}=0.0064$ und $^{239}$Pu mit $\beta_{Pu-239}=0.0021$ und $\Lambda=10^{-4}$s.
\end{center}

\subsection*{Berechnung der Produktionswahrscheinlichkeit $\mathcal{P}$}

\subsubsection*{Mikroskopische Wirkungsquerschnitte}
Die mit dem griechischen Symbol $\sigma$ bezeichneten mikroskopischen Wirkungsquerschnitte stellen einen zentralen
Begriff der Kernphysik und der Kerntechnik dar, welcher mit der praktischen Berechnung von Neutronenreaktionsraten
in unmittelbar Weise verkn\"upft ist. Aus der Sicht eines einzelnen Neutrons kann ein Reaktionspartner wie beispielsweise ein
$^{235}$U-Kern als eine Zielscheibe aufgefasst werden, deren Fl\"achennormale sich jeweils parallel zur Bewegungsrichtung
des Neutrons auftut. Die Zielscheibenfl\"ache l\"asst sich weiter in diverse Bereiche unterschiedlicher Gr\"osse aufgeteilt denken.
Trifft ein Neutron auf einen dieser Bereiche auf, so l\"ost es eine mit diesem Bereich assoziierte Reaktion mit dem Zielkern aus.
\\

\noindent Das in der Abbildung 7 dargestellte Sinnbild zeigt solche Fl\"achenbereiche f\"ur die wichtigsten Reaktionen mit spaltbaren
Kernen wie dem $^{235}$U-Kern als konzentrische Ringe dargestellt. Der zentrale Kreis steht sinngem\"ass f\"ur den
Neutroneneinfangquerschnitt $\sigma_\gamma$, welcher f\"ur Neutronen mit einer f\"ur den thermischen Bereich
typischen Energie von $k_B \cdot 293$K $= 0.025$eV etwa $95$b $=95 \cdot 10^{-24}$cm$^{2}$ betr\"agt. Die in der Teilchenphysik
gebr\"auchliche Fl\"acheneinheit von $10^{-24}$cm$^2$ ist das Barn (englisch f\"ur Scheune) mit dem Einheitenzeichen b.
Trifft ein Neutron in diesen Bereich der Zielscheibe, so bildet es mit dem $^{235}$U-Kern einen langlebigen, f\"ur die
Kettenreaktion nutzlosen $^{236}$U-Kern, der lediglich zum radioaktiven Abfallinventar eines Reaktors beitr\"agt.
Der durch den Neutronentreffer angeregte $^{236}$U-Kern emittiert seine \"ubersch\"ussige Energie
in Form von $\gamma$-Quanten, was die Indizierung des Wirkungsquerschnitts erkl\"art.
\\

\noindent Ein Treffer des Spaltwirkungsquerschnitts $\sigma_f$ l\"ost aber eine Kernspaltung aus. F\"ur thermische Neutronen
der kinetischen Energie $0.025$eV ist $\sigma_f = 586$b; allerdings ist der Spaltquerschnitt wie der Neutroneneinfangquerschnitt
stark von der Neutronenenergie abh\"angig und sinkt im MeV-Bereich auf die Gr\"ossenordnung von einem Barn. \\

\noindent Mit einer im thermischen Bereich bedeutend kleineren Wahrscheinlichkeit wird ein Neutron bei einer Kollision mit
einem $^{235}$U-Kern lediglich umgelenkt und streut elastisch ($\sigma_n$) oder inelastisch ($\sigma_{n^*}$), wobei es bei
der inelastischen Streuung den Kern  in einem angeregten Zustand hinterl\"asst.
\\

\noindent $\sigma_{Rest}$ in Abbildung 7 steht f\"ur
s\"amtliche \"ubrigen Neutronenreaktionen wie zum Beispiel dem Herausschlagen eines einzelnen Protons
bei gleichzeitiger Einverleibung des Neutrons im Restkern. Die Untersuchung solcher Reaktionen ist ein interessantes
Thema f\"ur sich,  f\"ur unsere Betrachtungen sind sie aber von geringer Bedeutung.
\\

\noindent Tats\"achlich ist die oben geschilderte Darstellung der Wirkungsquerschnitte k\"unstlich und entspricht nicht der Realit\"at,
die durch quantenmechanische Begriffe besser angen\"ahert werden kann. Dennoch lassen sich Wirkungsquerschnitte als
physikalische Beobachtungsgr\"ossen rigoros definieren und sind
im oben dargestellten Rahmen sinnvoll f\"ur praktische Berechnungen verwendbar.

\begin{center}
\includegraphics[width=11.0cm,angle=270]{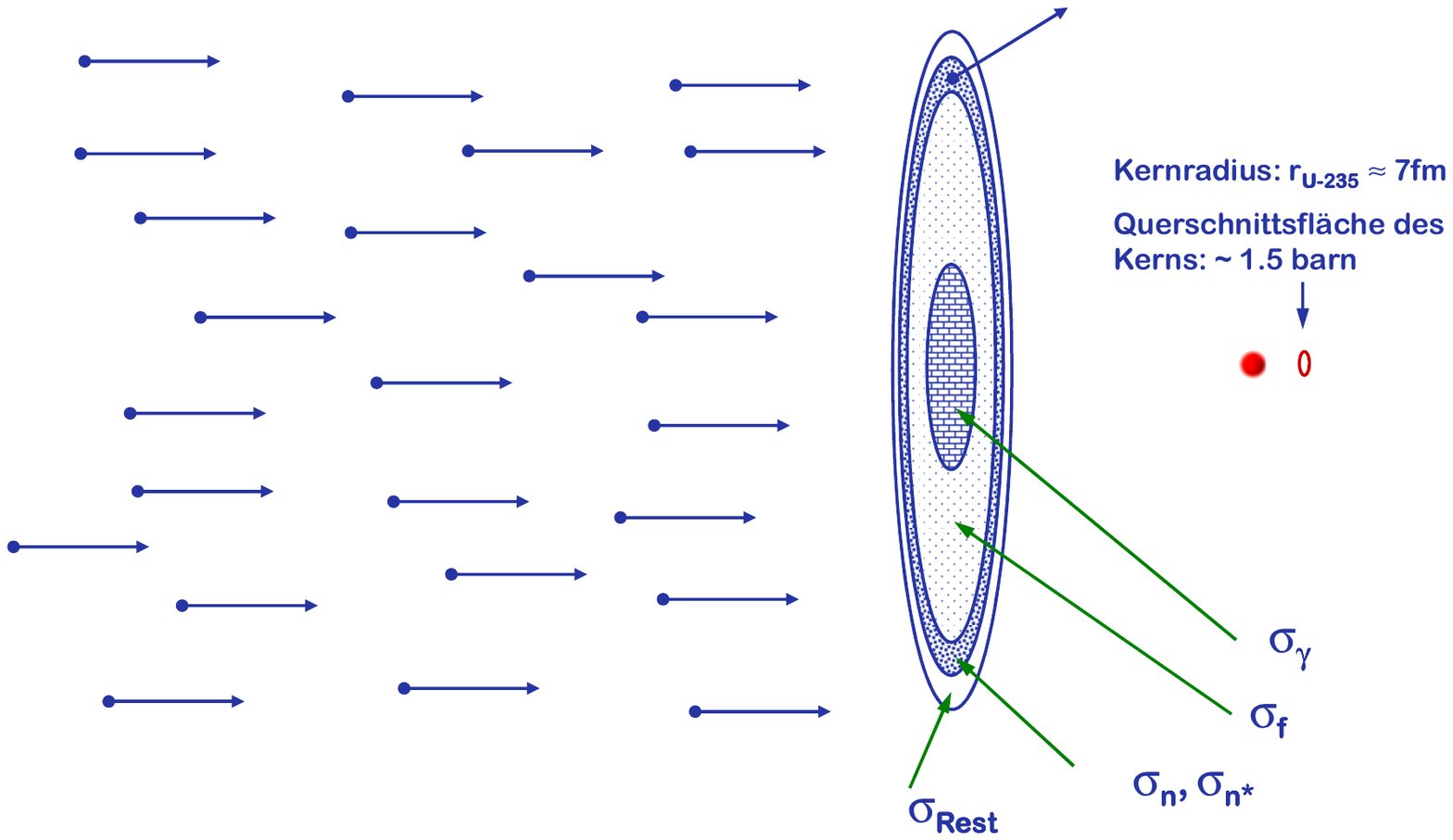}\\
\vspace{-0.5cm} Figur 7: Sinnbild f\"ur die relevanten Neutronen-Wirkungsquerschnitte eines $^{235}$U-Kerns.
\end{center}

\subsubsection*{Die Produktionswahrscheinlichkeit $\mathcal{P}$}

F\"ur das nun folgende Gedankenexperiment wollen wir als grobe N\"aherung annehmen, dass der Reaktorkern
homogen mit $^{235}$U als Spaltstoff belegt ist. Weiter soll auch die thermische Neutronenflussdichte im Reaktor in guter N\"aherung
mit einem effektiven Durchschnittwert $\Phi_{th}$ repr\"asentiert werden k\"onnen. Dies ist keine absolut gerechtfertigte Annahme, da die
Neutronenflussdichte in einem homogen beladenen Reaktorkern im Normalfall am Reaktorrand auf Grund der Leckage erniedrigt ist.
Dennoch stellt diese Tatsache f\"ur die folgenden prinzipiellen \"Uberlegungen keine wesentliche Beeintr\"achtigung dar.
Wir stellen uns also vor, dass sich s\"amtliche
Neutronen in einem Reaktor f\"ur eine sehr kurze Zeit und mit betragsm\"assig unver\"anderter Geschwindigkeit
in dieselbe Richtung bewegen. Es herrscht dann im Reaktor eine thermische Neutronenstromdichte
\begin{equation}
I_{th} = n_{th}^\Box \cdot \bar{v} \, ,
\end{equation}
wobei $\bar{v}$ die mittlere Neutronengeschwindigkeit der thermischen Neutronen ist.
Die thermische Neutronendichte l\"asst sich schreiben als
\begin{equation}
n_{th}^\Box = n/V \, ,
\end{equation}
wobei $V$ das Gesamtvolumen des Reaktorkerns bezeichnet.
$I_{th} \cdot \Sigma$ beschreibt die Anzahl Neutronen, die pro Zeiteinheit eine zur Bewegungsrichtung quer stehende
Fl\"ache $\Sigma$ durchdringen. Bezeichnen wir die Anzahl der $^{235}$U-Kerne im Reaktorkern mit $N_{U-235}$,
so spannen aber diese Kerne eine Gesamtfl\"ache $N_{U-235} \cdot \sigma_f$ f\"ur Kernspaltungen auf.

\begin{center}
\vskip -4.5cm
\includegraphics[width=11.0cm,angle=270]{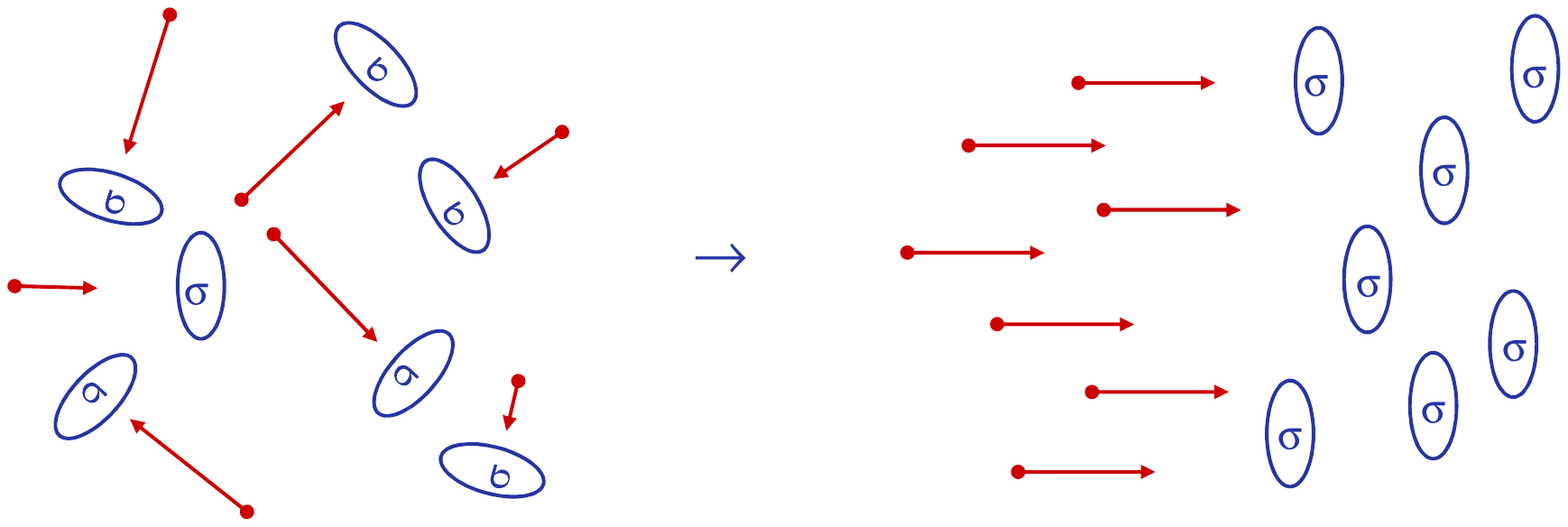}\\
\vspace{-1.5cm} Figur 8: Sinnbild f\"ur die begriffliche Unterscheidung einer Neutronenstromdichte (rechts) und einer
Neutronenflussdichte (links).
\end{center}

\noindent Wie in Abbildung 8 sinngem\"ass dargestellt spielt es f\"ur die Neutronen keine Rolle, ob sie sich in einem
homogenen Reaktor gerichtet als Strom oder ungerichtet als Neutronenfluss bewegen; jeder $^{235}$U-Kern
pr\"asentiert sich einem herannahenden Neutron als Zielscheibe mit den entsprechenden Wirkungsquerschnitten.
Wir k\"onnen bei der Berechnung von Reaktionsraten den Strom $I_{th}$ durch die thermische
Neutronenflussdichte $\Phi_{th}$ ersetzen,
die Kernspaltungrate $R_f$ im Reaktorkern ist also gegeben durch die Anzahl der Neutronen, welche die Gesamtfl\"ache
$\Sigma = N_{U-235} \cdot \sigma_f$ pro Zeiteinheit durchdringen
\begin{equation}
R_f = \Sigma \cdot \Phi_{th} = N_{U-235} \cdot \sigma_f \cdot n_{th}^\Box \cdot \bar {v} =
 N_{U-235}^\Box \cdot \sigma_f \cdot n_{th}\cdot \bar {v}\, ,
\end{equation}
wobei in der obigen Formel die Neutronendichte zur Neutronenzahl und die Spaltkernzahl zur Spaltkerndichte
umgeschrieben wurde. Da bei einer Kernspaltung durchschnittlich $\nu \simeq 2.5$ schnelle Neutronen entstehen,
aus welchen wiederum durchschnittlich $W_{th} \cdot \nu$ thermische Neutronen entstehen, ist die thermische Neutronenproduktionsrate
gegeben durch
\begin{equation}
P= \mathcal{P} \cdot n = W_{th} \cdot \nu \cdot \bar{v} \cdot N_{U-235}^\Box \cdot \sigma_f \cdot n_{th} \, ,
\end{equation}
d.h. es gilt f\"ur die Neutronenproduktionswahrscheinlichkeit
\begin{equation}
\mathcal{P}= W_{th} \cdot \nu \cdot \bar{v} \cdot \Sigma_f \, ,
\end{equation}
wobei der in der Kerntechnik verwendete
sogenannte makroskopische Spaltwirkungsquerschnitt $\Sigma_f = N_{U-235}^\Box \cdot \sigma_f$ eingef\"uhrt wurde
(Spalt(wirkungs)querschnitt pro Volumeneinheit).
\\

\noindent Es bleibt zu bemerken, dass  sich der thermische $^{235}$U-Spaltquerschnitt etwa invers proportional zur Neutronengeschwindigkeit
verh\"alt, weswegen f\"ur Rechnungen bei h\"oheren Temperaturen in guter N\"aherung die Wirkungsquerschnitte bei
Raumtemperatur und die sogenannte Westcott-Geschwindigkeit der Neutronen bei Raumtemperatur $\bar{v} \simeq
2.2 \cdot 10^5$cm$\cdot$s$^{-1}$ verwendet werden k\"onnen. Der Leser ist dazu eingeladen, die Neutronenproduktionswahrscheinlichkeit
$\mathcal{P}$ f\"ur einen 3GW-Siedewasserreaktor mit zu 3\% angereichertem Spaltstoff der Uran-Gesamtmasse $100$t und einem
Reaktorvolumen von $50$m$^3$ zu berechnen.

\end{document}